\documentclass[10pt,aps,prd,twocolumn,superscriptaddress,nofootinbib,nobibnotes,longbibliography,floatfix]{revtex4-2}

\PassOptionsToPackage{dvipsnames}{xcolor}
\usepackage{orcidlink}
\usepackage{aasmacros}
\usepackage[utf8]{inputenc}
\usepackage[T1]{fontenc}

\usepackage{mathtools,amsmath,amssymb,amsfonts}
\usepackage{bm}
\usepackage{pifont}
\usepackage{lmodern}
\usepackage{mathrsfs}
\usepackage[normalem]{ulem}
\usepackage{bigints}

\usepackage[dvipsnames]{xcolor}
\usepackage{graphicx} %
\usepackage{float}

\usepackage{hyperref}
\hypersetup{colorlinks=true, 
            citecolor=MidnightBlue,
            linkcolor=MidnightBlue, 
            urlcolor=MidnightBlue, 
            linktocpage=true}
\usepackage[capitalize]{cleveref}
\crefname{section}{Sec.}{Secs.}
\crefname{appendix}{App.}{Apps.}
\usepackage{array}
\usepackage[draft,defaultcolor=olive]{changes}

\usepackage{caption}
\definechangesauthor{RB}

\begin{document}

\title{
Muffled Murmurs: Environmental effects in the LISA stochastic signal from stellar-mass black hole binaries
}%
\author{Ran Chen}
\email{ranchen@pmo.ac.cn}
\affiliation{Key Laboratory of Dark Matter and Space Astronomy, Purple Mountain Observatory, Chinese Academy of Sciences, Nanjing 210033, People's Republic of China}
\affiliation{School of Astronomy and Space Sciences, University of Science and Technology of China, Hefei 230026, People’s Republic of China}
\affiliation{SISSA, Via Bonomea 265, 34136 Trieste, Italy}

\author{Rohit S. Chandramouli}
\email{rchandra@sissa.it}
\affiliation{SISSA, Via Bonomea 265, 34136 Trieste, Italy}
\affiliation{INFN Sezione di Trieste, Trieste, Italy}
\affiliation{IFPU - Institute for Fundamental Physics of the Universe, Via Beirut 2, 34014 Trieste, Italy}

\author{Federico Pozzoli}
\affiliation{Dipartimento di Scienza e Alta Tecnologia, Università dell'Insubria, via Valleggio 11, I-22100 Como, Italy}
\affiliation{INFN, Sezione di Milano-Bicocca, Piazza della Scienza 3, 20126 Milano, Italy}
\affiliation{Como Lake Centre for AstroPhysics (CLAP), DiSAT, Università dell'Insubria, via Valleggio 11, 22100 Como, Italy}

\author{Riccardo Buscicchio}
\affiliation{INFN, Sezione di Milano-Bicocca, Piazza della Scienza 3, 20126 Milano, Italy}
\affiliation{Dipartimento di Fisica ``G. Occhialini'', Università degli Studi di Milano-Bicocca, Piazza della Scienza 3, 20126 Milano, Italy}
\affiliation{Institute for Gravitational Wave Astronomy \& School of Physics and Astronomy, University of Birmingham, Birmingham, B15 2TT, UK}

\author{Enrico Barausse}
\affiliation{SISSA, Via Bonomea 265, 34136 Trieste, Italy}
\affiliation{INFN Sezione di Trieste, Trieste, Italy}
\affiliation{IFPU - Institute for Fundamental Physics of the Universe, Via Beirut 2, 34014 Trieste, Italy}

\date{\today}%

\begin{abstract}
    The population of unresolved stellar-mass black hole binaries (sBBHs) is expected to produce a stochastic gravitational-wave background (SGWB) potentially detectable by the Laser Interferometer Space Antenna (LISA).
    In this work, we compute the imprint of astrophysical environmental effects—such as gas dynamical friction and accretion—on this background.
    Using the sBBHs population constraints obtained by the LIGO--Virgo--Kagra collaboration, we compute the expected SGWB and develop a phenomenological parametric model
    that can accurately capture the effect of dynamical friction and accretion.
    Using our model,
    we perform Bayesian inference on simulated signals to assess the detectability of these environmental effects. 
    We find that even for large injected values of the Eddington ratio, the effect of accretion in the SGWB is undetectable by LISA. 
    However, LISA will be able to constrain the effect of dynamical friction with an upper bound on the gas density of 
    $\rho \lesssim 7.6 \times 10^{-10} \mathrm{g \, cm^{-3}}$, thus probing the sBBH environment forming in typical thin accretion disks around Active Galactic Nuclei (AGNs). 
    For injected densities of $\rho \sim 10^{-10}-10^{-9} \mathrm{g} \, \mathrm{cm}^{-3}$, dynamical friction effects can be well measured and clearly distinguished from vacuum, with Bayes factors reaching up to $\sim$ 60, even when the Galactic foreground is included.
\end{abstract}

\maketitle

\section{Introduction}
The detection of compact binary coalescences by the LIGO-Virgo-KAGRA (LVK) collaboration offers valuable insight into the stellar-mass black hole binaries (sBBHs) evolution, formation channels, and population properties~\cite{LIGOScientific:2016aoc,LIGOScientific:2020ibl,LIGOScientific:2018mvr,KAGRA:2021vkt,KAGRA:2021duu}.
GW190521 is an exceptional gravitational-wave (GW) event, with the primary black-hole (BH) mass residing in the mass gap predicted by pair-instability supernova
theory, therefore challenging current astrophysical formation scenarios~\cite{LIGOScientific:2020iuh,LIGOScientific:2020ufj}.
One possible explanation is that this event occurred within the gaseous disk of an active galactic nucleus (AGN), where mass segregation and dynamical friction drive the migration of black holes near the disk’s center, enhancing merger rates and facilitating mass growth through repeated mergers and sustained accretion~\cite{Levin:2006uc,Bartos:2016dgn,Tagawa:2019osr}.
Therefore, probing environmental effects through gravitational-wave observations is essential for advancing our understanding of astrophysical processes.

The Laser Interferometer Space Antenna (LISA)~\cite{amaro2017laser}, a planned space-based gravitational-wave observatory operating in the mHz frequency band, is well suited for detecting the dynamical signatures induced by astrophysical environments~\cite{Barausse:2020rsu,LISA:2022yao,2023arXiv231101300L,LISA:2022kgy}.
A primary reason for this is that environmental effects are typically more significant earlier in the inspiral.
In the LISA band, extreme mass ratio inspirals and intermediate mass black hole binaries are typical probes of the strong dynamics induced by the astrophysical environment~\cite{Barausse:2007dy,Yunes:2010sm,Gair:2010iv,Kocsis:2011dr,Barausse:2014tra,Deme:2020ewx,Caputo:2020irr,Chandramouli:2021kts,Speri:2022upm,Copparoni:2025jhq}.
Yet, the resolvable sBBHs that form in gas-rich environments, e.g. in the disks of AGNs, are also potentially sensitive to environmental effects~\cite{Toubiana:2020drf,Caputo:2020irr,Sberna:2022qbn,Kuntz:2022juv}.
However, the largest majority of sBBHs will not be detectable in the LISA band, resulting in the build-up of a stochastic gravitational-wave background (SGWB)~\cite{2025JCAP...01..084B,Lehoucq:2023zlt,Babak:2023lro}.
In general, a SGWB is a superposition of GW signals from an unresolved population of sources~\cite{Allen:1997ad,phinney2001practical,Maggiore:2007ulw}.
Computing the SGWB from unresolved sBBHs in the LISA band has been recently explored with the assumption that the the binaries are in vacuum~\cite{Babak:2023lro,Chen:2018rzo,Lewicki:2021kmu,Perigois:2020ymr}.
Since the amplitude of the vacuum SGWB from sBBHs may be measurable at the percent level \cite{Babak:2023lro}, we extend previous vacuum-based studies by investigating how environmental effects could imprint observable deviations.

Environmental effects typically induce additional energy dissipation that can dominate over the GW flux at low frequencies~\cite{Kocsis:2011dr,Barausse:2014tra}, thereby suppressing the SGWB relative to the vacuum case.
A similar scenario has been proposed to explain pulsar timing array (PTA) measurements~\cite{NANOGrav:2023gor, NANOGrav:2023hvm, NANOGrav:2023hfp, EPTA:2023fyk, EPTA:2023gyr, Xu:2023wog}, where the observed SGWB may originate from a population of supermassive binary black holes influenced by environmental effects~\cite{Chen:2016zyo, Bonetti:2017lnj, Aghaie:2023lan, Ghoshal:2023fhh}.
The main environmental effects expected to affect sBBHs in gas-rich environments are
\emph{dynamical friction} and \emph{accretion}~\cite{Barausse:2014tra,Caputo:2020irr,Toubiana:2020drf,Sberna:2022qbn} (see also~\cite{Kocsis:2011dr} for an overview of other environmental effects).
Dynamical friction is the result of the gravitational interaction of each black hole with the density wake produced by its motion through a fluid, collisionless~\cite{chandrasekhar_DF1} or collisional~\cite{Kim:2007zb,Barausse:2007ph,Barausse:2014tra}. Accretion
affects the binary because the infalling gas carries energy and momentum, which are transferred to the black hole and change its mass and momentum~\cite{Barausse:2007dy,Caputo:2020irr,Toubiana:2020drf}.
Thus, we focus on the imprints of dynamical friction and accretion on the SGWB of sBBHs.

Treating dynamical friction and accretion effects as perturbations on the Keplerian orbital motion, we derive their effect on the spectrum of the stochastic
background from a population of sBBHs consistent with the observational constraints from LVK's third observing run (O3)~\cite{KAGRA:2021kbb,KAGRA:2021duu}. 
We construct phenomenological parametric models
for the SGWB with environmental effects, which can be readily used 
with the Bayesian tools developed in~\cite{Pozzoli:2024,Pozzoli:2024hkt} to
quantify the detectability of accretion and dynamical friction with LISA observations.

We find that the effect of gas accretion is not detectable in the LISA stochastic background,
if the accretion rate is Eddington-limited, with the Eddington rate $f_{\rm Edd} \lesssim 10$. 
Meanwhile, dynamical friction
from gas densities comparable to those expected in AGN disks would yield a measurable effect on the background's spectrum. 
In more detail, LISA's observations of the stochastic background will probe gas densities $\rho \gtrsim \mbox{a few}\, \times 10^{-10} \mathrm{g \, cm^{-3}}$, with densities
larger than $ 10^{-9} \mathrm{g} \, \mathrm{cm}^{-3}$ detectable with large Bayes factors,
even when accounting for the effect of the Galactic white-dwarf foreground.
We also find that a sub-population of sBBHs undergoing dynamical friction in typical AGN disk gas densities can be probed.
Thus, this may offer potential insight into the formation channels of sBBHs.

Since the SGWB can also be affected by  modifications/extensions of General Relativity (GR)~\cite{Yunes:2025xwp,Maselli:2016ekw,Callister:2023tws,Cannizzaro:2023mgc,Chen:2024pcn}, we discuss how those effects can be mapped into our phenomenological analytic model for the spectrum.
For the specific case of a time-dependent Newton's constant~\cite{Yunes:2009bv}, we show that the SGWB produced by sBBHs in the LISA band can independently constrain $|\dot{G}/G| \lesssim 10^{-4} \mathrm{yr}^{-1}$, which is comparable to the projected bounds from quasi-monochromatic LISA sources~\cite{Barbieri:2022zge}.

This paper is organized as follows.
In Sec.~\ref{sec:SGWB_vac}, we review the SGWB from sBBHs
and present the results in the vacuum case.
For each dynamical friction and gas accretion model considered, we derive parametric phenomenological models for the energy spectrum and the resulting SGWB in Sec.~\ref{sec:imprint}.
In Sec.~\ref{sec:Bayesian_PE} we discuss the detectability and parameter estimation of environmental effects using our phenomenological models. 
We discuss how our phenomenological models can be used to generically probe additional dissipative channels in Sec.~\ref{subsec:agnostic}, and present our conclusions in Sec.~\ref{sec:conclusion}.
Throughout the paper, we use geometrized units in which $G=c=1$, unless otherwise specified.
We denote the component masses of the binary system by $m_1$ and $m_2$, with the total mass $m=m_1+m_2$, the reduced mass $\mu=m_1 m_2 / m$, the symmetric mass ratio $\eta = \mu/m$, and the chirp mass $\mathcal{M}=\mu^{3/5} m^{2/5}$.

\section{Stochastic gravitational-wave background from stellar binaries in vacuum}{\label{sec:SGWB_vac}}
A Gaussian, isotropic, unpolarized, and SGWB is fully characterized by its spectral energy density, $\Omega_{\mathrm{GW}}(f)$, given by~\cite{Allen:1997ad,phinney2001practical,KAGRA:2021kbb}
\begin{equation}
\Omega_{\mathrm{GW}}(f)=\frac{1}{\rho_c} \frac{d \rho_{\mathrm{GW}}}{d \ln f},
\end{equation}
where $\rho_{\mathrm{GW}}$ denotes the gravitational wave energy density, while the present critical energy density is given by $\rho_{c}=3 H_{0}^{2}/(8\pi)$.
Further, the spectral energy density $\Omega_{\mathrm{GW}}(f)$
from coalescing binary systems can be equivalently expressed as~\cite{Allen:1997ad,phinney2001practical,KAGRA:2021kbb}

\begin{equation}
\label{eqn:Omega_GW_calculation}
\begin{aligned}
\Omega_{\mathrm{GW}}(f)=\frac{f}{\rho_{c} H_0} &\iint dz d\boldsymbol{\phi}\frac{R_{\mathrm{GW}}(z)}{(1+z) \mathcal{E}(z)}  p(\boldsymbol{\phi}) \left. \frac{dE_{\rm GW}}{df_{s}} (\boldsymbol{\phi}) \right|_{f_s},
\end{aligned}
\end{equation}
where $\frac{dE_{\rm GW}}{df_s}(\boldsymbol{\phi})|_{f_s}$ is the source-frame energy spectrum radiated by a single source, evaluated at the source GW frequency $f_s = f(1+z)$ with $f$ being the detector frame
GW frequency. 
In~\cref{eqn:Omega_GW_calculation}, the integration is performed over the distribution $p(\boldsymbol{\phi})$ of source parameters $\boldsymbol{\phi}$ (e.g. masses, spins, etc.).
The quantity $R_{\mathrm{GW}}(z)$ is the comoving merger rate density of GW sources measured in the source frame, and $\mathcal{E}(z) = \sqrt{\Omega_m(1+z)^3+\Omega_{\Lambda}}$.
We adopt the result of Planck18~\cite{Planck:2018vyg} for the value of cosmology parameters.

For a binary with masses $m_1, m_2$, to leading order in the post-Newtonian (PN) expansion\footnote{In PN theory one expands in the weak-field, slow-motion limit. An $n$PN term scales as $v^{2n}$ relative to the leading contribution~\cite{Maggiore:2007ulw}, with $v = (\pi m f_s)^{1/3}$ being the orbital velocity.}
, the energy spectrum carried by gravitational waves emitted by a binary at a frequency $f_s$ is given by~\cite{phinney2001practical,Maggiore:2007ulw} 
\begin{align}
\dfrac{dE_{\rm GW}}{df_s} \equiv \dfrac{\dot{E}_{\rm GW}}{\dot{f_s}} = \frac{ \eta m^{5/3} \pi^{2/3}}{3 f_s^{1/3}},
\label{eqn:dEdf_vac}
\end{align}
where the chirp rate $\dot{f_s}$ is given by
\begin{align}
\label{eqn:fdot_vac}
    \dfrac{df_s}{dt} = \dfrac{96}{5} f_s^{11/3} m^{5/3} \pi^{8/3} \eta. 
\end{align}
From~\cref{eqn:Omega_GW_calculation,eqn:dEdf_vac}, the frequency-independent contribution to the SGWB can be absorbed into an overall amplitude $A_{\rm vac}$, yielding
\begin{align}
 \Omega_{\rm GW}(f) = A_{\rm vac} f^{2/3}. 
\end{align}

We adopt a standard astrophysical prescription for the merger rate and source population~\cite{KAGRA:2021duu,Madau:2014bja,Langer:2005hu,Callister:2023tws,Talbot:2018cva,Regimbau:2011rp}.  
The redshift-dependent merger rate is assumed to follow the cosmic star formation rate~\cite{Madau:2014bja}, weighted by metallicity~\cite{Langer:2005hu} and convolved with a distribution of time delays~\cite{Callister:2023tws}.  
The mass distribution follows the Power Law + Peak model~\cite{Talbot:2018cva} with negligible spins, consistent with LIGO-Virgo observations~\cite{LIGOScientific:2020kqk,Miller:2020zox}. 
With these models and the posterior distributions of their parameters inferred from Refs.\cite{KAGRA:2021kbb,KAGRA:2021duu}, we numerically evaluate~\cref{eqn:Omega_GW_calculation} via Monte Carlo integration to generate posterior predictions for the SGWB.
The median SGWB posterior prediction and the corresponding model parameters are adopted as fiducial values.
A comparison with the O3-based forecast is shown in Appendix~\ref{app:pop_model}, demonstrating consistency across the relevant frequency range.
Further, we note that one can alternatively use the chirp rate $\dot{f_s}$ to characterize the binary evolution and compute the number of systems within a given GW frequency bin of interest~\cite{Sesana:2008mz}.
In our approach, however, the inclusion of additional environmental dissipative channel modifies the spectral shape through its impact on the energy spectrum, while the overall normalization remains fixed by the merger rate informed by LVK population constraints, rather than by renormalizing the number of sources per frequency bin.

\section{Environmental-effects imprint on the SGWB}\label{sec:imprint}
We consider sBBHs embedded in a gaseous environment, so that the binary undergoes orbital evolution due to both the environment and the back-reaction from gravitational-wave emission.
Examples of such systems include sBBHs that form in the accretion disk of AGNs.
We focus on the imprint of accretion disk effects, primarily \emph{dynamical friction} and \emph{gas accretion}, on the SGWB from sBBHs. 
In the following, we analyze separately the effect of dynamical friction and accretion on the SGWB spectrum, and develop phenomenological analytic models for both.

\subsection{Dynamical friction}\label{subsec:eff_DF}
Density wakes are produced due to the motion of each black hole through the accretion disk.
Consequently, when the disk density is greater in the region trailing the black hole (compared to the region leading it), there is an opposing force to the black hole's motion, which is the cause of dynamical friction~\cite{chandrasekhar_DF1,Ostriker:1998fa,Kim:2007zb,Barausse:2007ph,Barausse:2014pra,Barausse:2014tra}. 
Such a force causes the binary black hole system to transfer binding energy to the gas.
Thus, in addition to GW emission, energy dissipates through another channel. 
When such effect dominates over GW emission, a drop in the GW energy spectrum, hence in the SGWB, is expected.

To compute it, we choose the center-of-mass (CoM) frame and assume that the disk is co-moving with the CoM.
We model the dynamical friction using a Newtonian approximation, relativistic effects being negligible corrections at the frequencies of interest for this study.
The dynamical friction force on a black hole depends on its mass $m_A$, the local disk density $\rho$ and speed of sound $v_s$, their relative velocity $\vec{v}_A$, and the Coulomb logarithm $I_A$.
Doing so, the dynamical friction force $\vec{F}_{\rm DF,A}$ on the $A$-th body is expressed as~\cite{Kim:2007zb,Ostriker:1998fa,chandrasekhar_DF1}
\begin{align}
    \vec{F}_{\mathrm{DF},A} = -\dfrac{4 \pi \rho m_A^2}{v_A^2} I_A \hat{v}_A. \label{eqn:DF_force}
\end{align}
Further, we assume that each black hole is moving at highly supersonic speeds relative to the local sound speed of the disk, i.e. yielding a Mach number $\mathscr{M}_A \equiv v_A / v_s \gg 1$.
For a typical environment where the sBBH 
is embedded in the AGN disk of a supermassive black hole, 
for a wide range of systems, the CoM velocity is small compared to the stellar binary velocity~\cite{Toubiana:2020drf}.
For circular orbits, we only need the azimuthal component of the force, and for supersonic motion, the azimuthal Coulomb logarithm $I_A$ (based on~\cite{Kim:2007zb}) to leading order in $\mathscr{M}_A \gg 1$ (neglecting $\mathcal{O}(1/\mathscr{M}_A)$ corrections) can then be expressed as 
\begin{align}
    I_A = \ln \left( \dfrac{100 r_A}{11 \mathscr{M}_A r_{\mathrm{min},A} }  \right),
\end{align}
with $r_{\mathrm{min},A} = 2 m_A$~\cite{Toubiana:2020drf}.
Since $r_{\min,A}$ is effectively the smallest length scale in the system arising from the regularization of the dynamical friction force integrals~\cite{Kim:2007zb}, as long as $r_{\min,A} \ll r_A$, the precise parameterization of $r_{\min,A}$ is not important.
We checked that alternative choices of $r_{\min,A}$ (see for e.g.,~\cite{Kim:2007zb,Barausse:2007ph,Cardoso:2019rou}) yield negligible impact on our results.
We expand on the validity of our dynamical friction modeling in Appendix~\ref{app:validity}.

The resulting (outgoing) energy flux lost to dynamical friction is given by $\dot{E}_{\rm DF} = - \sum_A \vec{F}_{\mathrm{DF},A} \cdot \vec{v}_A $, which becomes
\begin{align}
 \dot{E}_{\rm DF} = \dfrac{4 \pi^{2/3} \rho \ m_1^2 m^{2/3}}{m_2 f_s^{1/3}} \ln \left( \dfrac{f_{1}^{*}}{f_s}\right) + (m_1 \leftrightarrow m_2),
\end{align}
where $f_{A}^{*} = 50 v_s / (11 \pi m_A)$ and the shorthand $(m_1 \leftrightarrow m_2)$ denotes the same term with $m_1$ and $m_2$ interchanged.
From the energy-balance law, the rate of change of binding energy $E_{\rm b}$ is given by $\dot{E}_{\rm b} = -\dot{E}_{\rm GW} - \dot{E}_{\rm DF}$, yielding an additional contribution to $\dot{f}$:
\begin{align}
\dfrac{df_s}{dt} \!= \!\left(\dfrac{df_s}{dt} \right)_{\rm vac}\!\!\!\!+\! \dfrac{12\rho}{m^3 \eta^2}\! \left[ m_1^3 \ln \left(\dfrac{f_{1}^{*}}{f_s} \right)\!+\!(m_1 \leftrightarrow m_2) \right],
\end{align}
where $(df_s/dt)_{\rm vac}$ is given by Eq.(\ref{eqn:fdot_vac}).
Thus, the GW energy spectrum is modified in the presence of dynamical friction and reads 
\begin{align}
\left( \frac{dE_{\rm GW}}{df_s} \right)_{\rm DF} &= \left( \frac{dE_{\rm GW}}{df_s} \right)_{\rm vac} \left\{ 1 + \frac{5 \rho f_s^{-11/3}}{8 \pi^{8/3} \eta^3 m^{14/3}} \right. \nonumber \\
& \times \left. \left[ m_1^3 \ln \left( \dfrac{f_{1}^{*}}{f_s}\right) + (m_1 \leftrightarrow m_2)\right] \right\}^{-1},    \label{eqn:dEdf_DF}
\end{align}
where $(dE_{\rm GW}/df_s)_{\rm vac}$ is given by~\cref{eqn:dEdf_vac}.
The energy flux due to dynamical friction is a -5.5PN relative to the GW flux, owing to the $f_s^{-11/3}$ scaling. 
Consequently, at low frequencies dynamical friction will dominate the orbital evolution and energy loss, and can deplete the SGWB.

At a critical turning point frequency $f_{\rm turn,DF}$, the flux due to dynamical friction becomes comparable to that of GW emission.
For $f_s<f_{\rm turn,DF}$, the amplitude of the GW spectrum will start to decrease as more energy flows through the dynamical friction channel than the GW emission channel.
We compute $f_{\rm turn,DF}$ from $|\dot{E}_{\rm DF}| = |\dot{E}_{\rm GW}|$, which amounts to setting the $\rho$-dependent term in the denominator of~\cref{eqn:dEdf_DF} to 1.

To gain  insight onto $f_{\rm turn,DF}$, let us note that typical astrophysical systems yield $f_{A,*} \sim 10^{-1}$Hz, much higher than the observed frequencies that we want to probe.
Observing that $f_{A,*} \gg f_{\rm turn,DF}$ and focusing on nearly equal mass systems, we obtain
\begin{align}
    \dfrac{f_{\rm turn}}{\left(\ln[1\rm Hz/f_{\rm turn}] \right)^{3/11}} & \approx (1.69 \times 10^{-3} \mathrm{Hz}) \left( \dfrac{m}{50 M_{\odot}} \right)^{\!\!-5/11} \nonumber \\
   & \times \left( \dfrac{\rho}{10^{-10} \mathrm{g} \, \mathrm{cm}^{-3}} \right)^{3/11} ,\label{eqn:f_turn}
\end{align}
where we have scaled the estimate for typical astrophysical thin disk densities of $\rho \sim 10^{-10} \mathrm{g} \, \mathrm{cm}^{-3}$ and typical stellar masses of $m \sim 50 M_{\odot}$.

\begin{figure}
    \centering
    \includegraphics[width=\columnwidth]{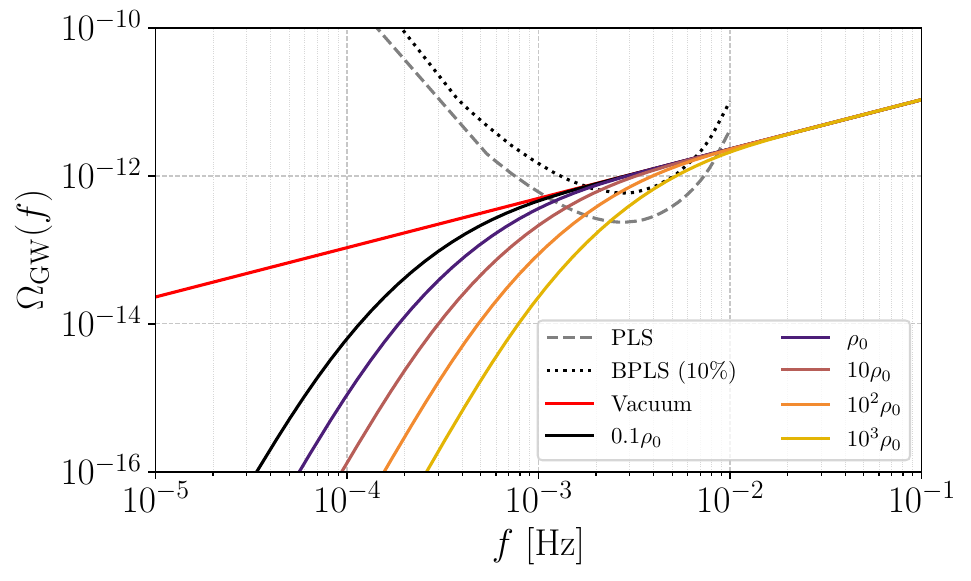}
    \caption{
The SGWB spectra $\Omega_{\mathrm{GW}}(f)$ for various disk densities $\rho$, with the normalization $\rho_0 = 10^{-10}\mathrm{g}\, \mathrm{cm}^{-3}$.
The solid lines correspond to different values of $\rho$, with the vacuum case corresponding to $\rho = 0$.
The gray dashed (black dotted) curves denotes the LISA power-law sensitivity (PLS) and Bayesian power-law sensitivity (BPLS), assuming a signal-to-noise ratio of 10, a Bayes factor threshold of 10, and 10\% noise level uncertainty.
    }
    \label{fig:DF_gwb}
\end{figure}

In~\cref{fig:DF_gwb}, we show the SGWB computed with~\cref{eqn:dEdf_DF}.
As for the vacuum case, we perform a three dimensional Monte Carlo population integral. 
SGWB are bracketed by a few reference $\rho$ values. 
As $\rho$ is increased, the turning point in the spectrum does increase as expected from~\cref{eqn:f_turn}.
Further, the spectral index of the SGWB asymptotes to a value of $13/3$ at low frequencies and $2/3$ at high frequencies, the latter being the vacuum result.
For typical disk densities of $\rho \sim 10^{-11} \mathrm{g} \, \mathrm{cm}^{-3}$--$\rho \sim 10^{-8} \mathrm{g} \, \mathrm{cm}^{-3}$, the stochastic signal typically lies above the power-law sensitivity (PLS) and Bayesian power-law sensitivity (BPLS) curves~\cite{Pozzoli:2024hkt} for $f \gtrsim 10^{-3}$Hz.
We note that both the PLS and BPLS curves assume power-law SGWB signals, with the signal-to-noise ratio (SNR) and Bayes factor quantifying signal detectability relative to noise, and are included here merely as references.
Additionally, for the typical disk densities considered here, we can expect that the dynamical friction effects are potentially measurable since the turning point occurs in the sensitive part of the LISA band.
Thus, the SGWB of sBBHs is potentially detectable and the effects of dynamical friction are also potentially measurable.
However, to rigorously determine the detectability of the signal and measurability of the parameters, we perform a detailed Bayesian analysis in~\cref{sec:Bayesian_PE}.

\subsection{Gas accretion}\label{subsec:eff_acc}
The gas from the surrounding disk will accrete onto the two black holes.
We model the accretion of the $A$-th body as $\dot{m}_{A,\rm Edd} = L_{A,\rm Edd}/\zeta$, where $L_{A,\rm Edd}$ is the Eddington luminosity and $\zeta$ is the radiative efficiency~\cite{Barausse:2014tra} (see also~\cite{Frank_King_Raine_2002} for a detailed review of accretion modeling).
We pick a conservative value for the radiative efficiency of $\zeta = 0.1$ and the resulting accretion rate is $\dot{m}_{A,\rm Edd} \simeq 2.2 \times 10^{-8} (m_A / M_{\odot}) M_{\odot} \mathrm{yr}^{-1}$~\cite{Barausse:2014tra,Caputo:2020irr}.
For simplicity, we parameterize the accretion rate of both black holes by the same Eddington ratio $f_{\rm Edd} \equiv \dot{m}_A / \dot{m}_{A,\rm Edd}$.
Doing so, the mass as a function of time reads~\cite{Caputo:2020irr}
\begin{equation}
\label{eqn:accreting mass}
m_A(t)=m_{A, 0} e^{f_{\mathrm{Edd}} t / \tau},
\end{equation}
where $\tau=4.5 \times 10^7 \mathrm{yr}$ is known as the Salpeter time scale, and $m_{A, 0}$ is the initial mass of the $A$-th black hole of binaries.

For a slowly accreting binary, the component masses evolve adiabatically, satisfying the condition $\dot{m}_A \ll m_A \omega_s /(2 \pi)$. 
In addition to the adiabatic mass increase of each body, the accreted material will additionally carry some angular momentum, which results in a hydrodynamic drag torque imparted on each body.
We parameterize this drag force as
\begin{align}
    \vec{F}_{\mathrm{Acc},A} =  - \xi \dot{m}_A \vec{v}_A, \label{eqn:drag_force}
\end{align}
where $\xi \sim \mathcal{O}(1)$ is the linear hydrodynamic drag coefficient that captures the effect of momentum transferred from the accreted material~\cite{Toubiana:2020drf,Barausse:2014tra,Gair:2010iv,Barausse:2007dy}.
Since angular momentum is an invariant under an adiabatic change in the masses~\cite{landau1976mechanics}, we can use angular momentum balance law to obtain the back-reaction $\dot{f}_s$ under gas accretion~\cite{Caputo:2020irr}. 
Doing so, we obtain
\begin{align}
    \left( \dfrac{df_s}{dt} \right)_{\rm Acc} = (5+3\xi)\dfrac{f_{\rm Edd}}{\tau} f_s. \label{eqn:dfdt_acc}
\end{align}
One could also obtain the same back-reaction using the energy-balance law but after accounting for the time-variation of the Hamiltonian under the adiabatic mass increase.
We explicitly show the equivalence of the two balance laws in Appendix~\ref{app:edd}, as an incorrect version of the energy balance law has been used in the literature~\cite{Yunes:2016jcc,Chamberlain:2017fjl,Barbieri:2022zge,Cannizzaro:2023mgc} to obtain the back-reaction for a time-dependent Hamiltonian.
The total back-reaction due to accretion and GW emission is then the sum of~\cref{eqn:dfdt_acc,eqn:fdot_vac}.
We also emphasize that~\cref{eqn:fdot_vac} is evaluated with the initial masses, i.e. we neglect contributions due to cross terms between accretion and radiation reaction because we effectively treat these effects to leading order in a multiple-timescale analysis~\cite{benderorszag}.

In this work, we do not consider relativistic corrections to the hydrodynamical drag (such as in~\cite{Barausse:2007dy,Gair:2010iv}) as (i) they are more important for extreme-mass-ratio inspirals than for comparable mass stellar binaries, and (ii) we are primarily interested in how accretion effects contribute as an additional energy dissipation channel, for which the non-relativistic treatment is sufficient. 
We comment further on the validity of the accretion modeling in Appendix~\ref{app:validity}.
Thus, using~\cref{eqn:dfdt_acc} together with~\cref{eqn:fdot_vac,eqn:dEdf_vac}, we obtain the energy spectrum of gravitational waves in the presence of accretion as
\begin{align}
\left( \frac{dE_{\rm GW}}{df_s} \right)_{\rm Acc} &= \left( \frac{dE_{\rm GW}}{df_s} \right)_{\rm vac} \nonumber\\
& \times \left[ 1 + \dfrac{5(f_{\mathrm{Edd}}/\tau)(5+3\xi)}{96\pi ^{8/3}m_0^{5/3}\eta }f_s^{-8/3}\right]^{-1},\label{eqn:dEdf_Acc}
\end{align}
where $(dE_{\rm GW}/df_s)_{\rm vac}$ is given by~\cref{eqn:dEdf_vac} and evaluated with the initial value of the masses.

\begin{figure}[!t]
    \centering
    \includegraphics[width=\columnwidth]{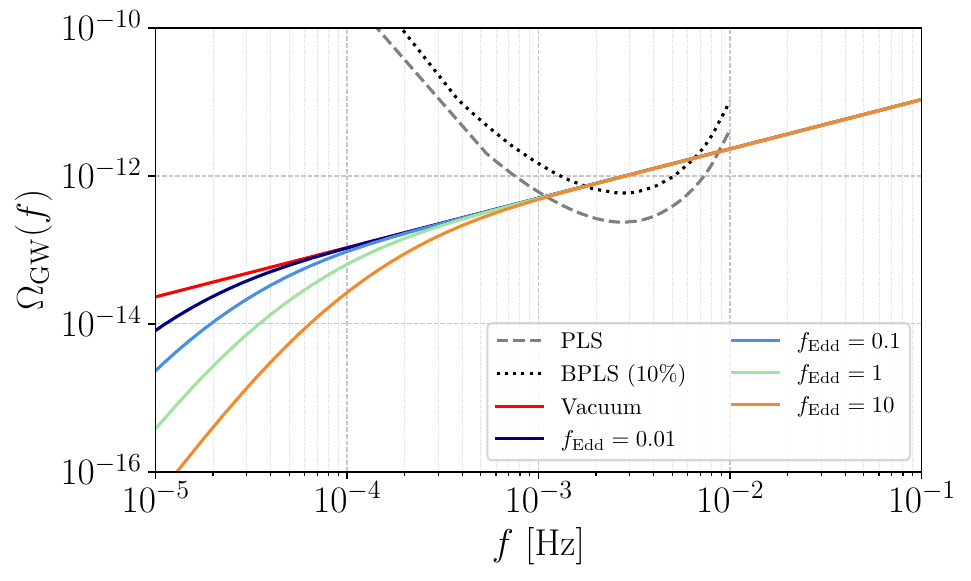}
    \caption{
The SGWB spectra $\Omega_{\mathrm{GW}}(f)$, driven by gas accretion, for a range of Eddington ratios $f_{\mathrm{Edd}}$.
The solid lines correspond to different values of $f_{\rm Edd}$, with the vacuum case corresponding to $f_{\rm Edd} = 0$.
The gray dashed and black dotted curves indicate the LISA power-law sensitivity (PLS) and Bayesian power-law sensitivity (BPLS) under the same assumptions made in~\cref{subsec:eff_DF}.
    }
    \label{fig:Acc_gwb}
\end{figure}

Similar to the case of dynamical friction, the stochastic GW signal will have a turning point due to additional energy dissipation in the presence of accretion.
At this turning point, the energy flux from gas accretion becomes comparable to that from gravitational-wave emission, i.e. $|\dot{E}_{\rm Acc}| = |\dot{E}_{\rm GW}|$, 
which is equivalent to setting the $f_{\mathrm{Edd}}$-dependent term in the denominator of Eq.~\eqref{eqn:dEdf_Acc} to 1.

We consider typical stellar binaries embedded in thin disks and estimate the turning point analytically as 
which 
\begin{align}
f_{\rm turn} &\approx 10^{-4}\mathrm{Hz}\,
\left(\frac{f_{\mathrm{edd}}}{1}\right)^{3/8}
\left(\frac{m}{50\,M_{\odot}}\right)^{-5/8} \nonumber\\
&\quad\times \left(\frac{5+3\xi}{8}\right)^{3/8}
\left(\frac{\tau}{4.5\times10^{7}\,\mathrm{yr}}\right)^{-3/8}\,. \label{eqn:f_turn_acc}
\end{align}

In addition, as the total mass increases, the turning point shifts toward lower frequencies.
For typical values of $f_{\rm Edd} \sim 1$, we find that $f_{\rm turn} \sim 10^{-4}$Hz, implying that the effects are significant only at the lower end of the LISA sensitivity band.
Only with much higher values of $f_{\rm Edd} \sim 10^3$, which may not be astrophysically realistic, we have $f_{\rm turn} \sim 10^{-3}$Hz, which is comparable to that of dynamical friction.
As expected, the effect of gas accretion occurs at -4PN order due to the $f_s^{-8/3}$ scaling. 
The key reason that gas accretion has an overall lower turning point than dynamical friction is its higher (less negative) PN order relative to GW emission.\footnote{Typically when a particular effect scales as $v^{-n}$ relative to GW emission, with $v = (\pi m f_s)^{1/3}$ and $n > 0$, the effect is stronger for larger $n$ due to $v \ll 1$.}. %

In~\cref{fig:Acc_gwb}, we show the SGWB computed with~\cref{eqn:dEdf_Acc} for different astrophysical values of $f_{\rm Edd}$, and we set $\xi = 1$ for simplicity.
As expected, the signal is suppressed at low frequencies due to the additional energy loss channel, and asymptotes to the vacuum SGWB at high frequencies. 
Although the stochastic signals shown in~\cref{fig:Acc_gwb} lie above the PLS and BPLS curves for the frequency range of $f\gtrsim 10^{-3}$Hz, suggesting that they are detectable, the turning points are outside the LISA SGWB sensitivity band for the range of astrophysical $f_{\rm Edd}$ values considered here.
Thus, we do not expect to clearly measure $f_{\rm Edd}$ with LISA and distinguish gas accretion from vacuum.
As with the case of dynamical friction, this requires a more rigorous Bayesian analysis, which we do in~\cref{sec:Bayesian_PE}.

\subsection{Phenomenological parametric models of stochastic signals containing environmental effects}\label{subsec:phenom_template}
The stochastic signals described in~\cref{subsec:eff_DF,subsec:eff_acc} are computationally expensive to evaluate for data analysis purposes.
In order to analyze the effects of dynamical friction and accretion, we develop ``ready-to-use'' phenomenological models.
As a first step, we construct a ``Rational Power-Law'' model $\Omega_{\rm RPL}$ given by 
\begin{align}
\Omega_{\mathrm{RPL}} = \frac{A_{\mathrm{vac}} f^{\gamma}}{1+A_{\mathrm{m}}\alpha f^{\beta}[\ln (f/1 {\rm Hz})]^\kappa}, \label{eqn:RPL_template}
\end{align}
where $A_{\rm vac}$ is the vacuum amplitude, $\gamma$ the vacuum spectral index, $\alpha$ is a phenomenological coefficient, $\{\beta,\kappa \}$ are spectral indices parameterizing each environmental effect, and $A_m$ controls the
fractional change in the SGWB amplitude due to the environment.
We obtain $A_{\rm vac}$ and $A_m$ using asymptotic matching at low and high frequencies. 
Specifically, we expand both $\Omega_{\rm RPL}$ and $\Omega_{\rm GW}$ to leading order for $f_s \ll f_{\rm turn}$ and for $f_s \gg f_{\rm turn}$ and solve for $A_{\rm vac}$ and $A_m$.
We provide details of the asymptotic matching in Appendix~\ref{app:template}, while values for $A_{\rm vac}$ and $A_m$ are listed in~\cref{tab:env_mapping_params}.

\begin{table}[!t]
    \centering
    \renewcommand{\arraystretch}{1.1}
    \begin{ruledtabular}
    \begin{tabular}{lcc}
        \textbf{Parameter} & \textbf{Dynamic Friction} & \textbf{Accretion} \\
        \hline
        $\alpha$ & $\rho$ & $(5 + 3\xi) \, f_{\mathrm{Edd}}/\tau$ \\
        $\beta$  & $-11/3$ & $-8/3$ \\
        $\kappa$ & $1$ & $0$ \\
        \hline
        \multicolumn{3}{l}{\textit{Asymptotic matching parameters}} \\
        $A_{\mathrm{vac}}$ & $4.97 \times 10^{-11}$ & $4.97 \times 10^{-11}$ \\
        $A_m$ & $-2.73 \times 10^{-7}$ & $4.35 \times 10^{2}$ \\
        \hline
        \multicolumn{3}{l}{\textit{Gaussian spectral correction parameters}} \\
        $A_g$ & $1.02 - 0.0262 \log_{10}(\alpha / \alpha_0)$ & $0.904$ \\
        $f_{\mathrm{peak}}$ & $0.0269\, (\alpha / \alpha_0)^{0.262}$ & $9.73\, (\alpha / \alpha_0)^{0.374}$ \\
        $\sigma$ & $0.222$ & $0.321$ \\        
    \end{tabular}
    \end{ruledtabular}
    \caption{The mapping parameters under different environmental effects. 
    The first three rows correspond to the environmental coefficient and respective spectral indices.
    The asymptotic matching and Gaussian correction parameters under different environmental effects are given in the next set of rows. 
    All coefficients are scaled to convenient SI units as follows: 
    $A_{\mathrm{vac}}$ in $\mathrm{Hz}^{-2/3}$; 
    $A_m$ in $\mathrm{kg}^{-1} \mathrm{m}^3 \mathrm{s}^{-11/3}$ (dynamical friction) and $\mathrm{s}^{-5/3}$ (accretion); 
    $f_{\mathrm{peak}}$ in Hz. 
    Here, $\alpha$ denotes the mapping parameter, normalized by $\alpha_0 = \mathrm{kg\, m^{-3}}$ (dynamical friction) and $\mathrm{s^{-1}}$ (accretion).
    }
    \label{tab:env_mapping_params}
\end{table}

At intermediate frequencies $f_s \sim f_{\rm turn}$, the asymptotic expansions break down and $\Omega_{\rm RPL}$ is not a sufficiently accurate approximation, as we explain in more detail in Appendix~\ref{app:template}. 
To improve the model accuracy at intermediate frequencies, we include a Gaussian correction to the RPL model.
The resulting ``Rational Power-Law + Peak'' model $\Omega_{\rm RPLP}$ performs well across all frequencies.
The model $\Omega_{\rm RPLP}$ is given by
\begin{align}
\begin{aligned}
\label{eqn:RPLP}
\Omega_{\mathrm{RPLP}} = \dfrac{\Omega_{\rm RPL}}{1+\mathcal{G}(f,\alpha)},
\end{aligned}
\end{align}
where the Gaussian correction $\mathcal{G}(f, \alpha)$ captures deviations near the turning point correction and reads
\begin{align}
\!\!\!\!\!\!\mathcal{G}(f,\alpha)\!=\!A_{g}(\alpha)\exp\!\left\lbrace-\frac{[\log_{10}(f/f_{\mathrm{peak}}(\alpha))]^2}{2 \sigma^2}\right\rbrace. \!\!
\end{align}
Here $A_g(\alpha)$, $f_{\mathrm{peak}}(\alpha)$ and $\sigma$  control the amplitude, central log-frequency, and width of the correction to $\Omega_{\rm RPL}$.
In~\cref{tab:env_mapping_params}, we show the mapping between  model parameters and the specific cases of dynamical friction and gas accretion.
We discuss the accuracy of our phenomenological models with respect to the ``accurate'' stochastic signals from~\cref{subsec:eff_DF,subsec:eff_acc} in Appendix~\ref{app:template}.

\section{Detection and parameter estimation of environmental effects} \label{sec:Bayesian_PE}
In this section, we first review how stochastic signals are analyzed in LISA data, then show results for how environmental effects (dynamical friction or accretion) can be measured using the phenomenological models developed in~\cref{subsec:phenom_template}.

\subsection{Detecting stochastic signals in LISA}\label{subsec:LISA_detection}
We model LISA data as linear, time-delayed combinations of six single-link measurements, known as time-delay interferometry (TDI) variables~\cite{Tinto:2021}. 
TDI variables are employed to suppress laser frequency noise, with different combinations applied depending on satellite orbits assumptions.~\cite{Cornish:2003tz,Shaddock:2003bc,Tinto:2003vj,Tinto:2021,Prince:2002hp}.
In this work, we use the uncorrelated TDI variables A, E, and T, suitable for a static, equal-arm of the LISA constellation~\cite{Prince:2002hp}.

Moreover, in our data model, we do not account for additional features beyond a stationary Gaussian spectrum such as anisotropy, non-Gaussianity, and non-stationarity. 
While these characteristics may help distinguish overlapping backgrounds, they also significantly increase the complexity of the analysis~\cite{Buscicchio:2024,Pozzoli:2025,2025PhRvD.111j3047P,Karnesis:2025,Criswell:2025}. 

Under these conditions, the stochastic signals in a TDI channel are described in Fourier domain as the superposition of the one-sided power spectral densities, since the signal and noise are treated as independent processes.

In each TDI channel, we infer on frequency domain data coarse grained over $N$ neighbouring points, resulting in $D = 5096$ segments, whose power spectral density (PSD) estimators for the $k-$th TDI variable are denoted with $\hat{S}_k(f_i)$, with $k=A,E,T$ and $i=1,\dots,D$.
Under the above assumptions, each $\hat{S}_k(f_i)$ is distributed according to a Gamma distribution~\cite{Appourchaux:2003}. 
The resulting log-likelihood reads
\begin{align}
\ln \mathcal{L}(\widehat{\boldsymbol{S}} \mid \theta) &=
\sum_{i = 1}^{D} \sum_{k} \left[
  -\ln \Gamma(N)
  - N \ln \left( \frac{S_{k}(f_i; \theta)}{N} \right) \right. \nonumber\\
  &\left. + (N - 1) \ln \widehat{S}_{k}(f_i)
  - \frac{N \, \widehat{S}_{k}(f_i)}{S_{k}(f_i; \theta)}
\right],
\end{align}
where $\theta$ are the model parameters and for compactness $\boldsymbol{\hat{S}}=(\hat{S}_A,\hat{S}_E,\hat{S}_T)$.

In the most general case, $S_{k}(f;\theta)$ includes contributions from instrumental noise, the Galactic foreground, and the superposition of multiple SGWBs, possibly accounting for environmental effects. 
Thus, in a given TDI channel, we have that
\begin{equation}
    S_{k}(f;\theta) = S_{k,\mathrm{GW}}(f;\theta) + S_{k,n}(f;\theta), 
\end{equation}
where $S_{k,\mathrm{GW}}(f;\theta)$ and $S_{k,n}(f;\theta)$ are signal and LISA noise PSD, respectively.
TDIs spectra  $S_{k,\mathrm{GW}}(f;\theta)$ can be conveniently recast as
\begin{equation}
    S_{k,\mathrm{GW}}(f;\theta) = R_{k}(f;\theta) \, S_\mathrm{GW}(f;\theta),
\end{equation}
where $R_{k}(f;\theta)$ is the LISA response function to an isotropic SGWB. 
Details on how to compute them can be found, e.g., in~\cite{Pozzoli:2024, Hartwig:2023, Baghi:2023qnq}.
By conventionally choosing $S_{\mathrm{GW}}(f;\theta)$ as the one sided GW  spectral density, we relate it to $\Omega_{\rm GW}$~\cite{Caprini:2018} as
\begin{equation}
    S_{\mathrm{GW}}(f;\theta) = \frac{3 H_0^2}{4 \pi^2 f^3} \Omega_{\mathrm{GW}}(f;\theta).
\end{equation}

We adopt a two-parameter model for the instrumental noise $S_{k,n}$, describing the spectral amplitudes of the test mass and optical metrology system noises, respectively, with parameters $\theta_n$~\cite{Quang:2023}.
The SGWB has instread contributions from both the population of Galactic white dwarfs and the sBBH, resulting in $S_{\mathrm{GW}} = S_{\mathrm{GW,Gal}}+S_{\mathrm{GW,sBBH}}$.
We adopt a phenomenological model for the Galactic SGWB contribution~\cite{Karnesisi:2021}, which reads
\begin{align}
    S_{\mathrm{GW,Gal}}(f; \theta_{\rm Gal}) & =\frac{A_\mathrm{Gal}}{2} f^{-7/3} \exp\left[-(f / f_1)^{\alpha_{\mathrm{Gal}}}\right] \nonumber \\
    & \times \left[1 + \tanh{((f_{\mathrm{kn}} - f)/f_2)}\right].
\end{align}
where $\theta_{\mathrm{Gal}} = \{\alpha_{\mathrm{Gal}}, A_{\mathrm{Gal}}, f_{\mathrm{kn}}, f_1, f_2\}$ are a set of suitable parameters capturing individual sources resolvability as a function of the mission duration.

We simulate data for $S_{\mathrm{GW,sBBH}}$ using $\Omega_{\rm GW, sBBH}$ models introduced in~\cref{sec:SGWB_vac,sec:imprint}.
Instead, for parameter estimation, we use the RPLP models introduced in~\cref{eqn:RPL_template,eqn:RPLP}, parameterized by $\theta_{\rm sBBH} = \{A_{\rm vac},\gamma,\alpha,\beta,\kappa, A_m\}$.
Thus, in the most general case, the inference parameter space is $\theta = \theta_n \cup \theta_{\rm Gal} \cup \theta_{\rm sBBH}$.

\subsection{Measuring environmental effects}\label{subsec:measurability}
In the following, we perform several injection-recovery studies to assess the measurability of the environmental effects.
Specifically, we assess separately the measurability of dynamical friction and gas accretion parameters.
To quantify the impact of the Galactic foreground, we perform a separate set of analyses where its parameters are assumed perfectly known.
Environmental effects are typically measurable when the recovered one-dimensional marginal posterior is well constrained relative to the prior.
To quantify the measurement precision for a parameter $\theta$, we quote the (symmetric, fractional) statistical uncertainty corresponding to the $90\%$ credible interval given by
\begin{equation}
    {\frac{\delta \theta}{\theta^{\rm inj}}} = \left| \dfrac{\theta^{95\%} - \theta^{5\%}}{2\theta^{\rm inj}} \right|,\label{eqn:stat_error}
\end{equation}
where $\theta^{\rm inj}$ denotes the true injected parameter value. Where suitable, we complement it with the SGWB SNR given by~\cite{Pozzoli:2024hkt}
\begin{equation}
    \mathrm{SNR} = \sqrt{T \sum_k 4\int_0^\infty df \left( \frac{S_{\rm k, GW}(f)}{S_{\rm k, n}(f)} \right)^2}, \label{eqn:snr}
\end{equation}
where $T$ is the observation time, set to $4\mathrm{yrs}$ for LISA.
We further assess the distinguishability from a vacuum signal by computing the Bayes factor between the non-vacuum and vacuum hypotheses (see~\cite{Thrane:2018qnx} for a GW specific review).
The log-Bayes factor in favor of the non-vacuum model over the vacuum one is given by
\begin{equation}
    \log_{10}\mathcal{B}^{\rm non-vac}_{\rm vac} = \log_{10}\mathcal{Z}_{\rm non-vac} - \log_{10}\mathcal{Z}_{\rm vac},
    \label{eq:bayes_factor}
\end{equation}
where the Bayesian evidence is given by
\begin{equation}
    \mathcal{Z}(\hat{\boldsymbol{S}}) = \displaystyle \int  d\theta\mathcal{L}(\hat{\boldsymbol{S}}|\theta)\Pi(\theta),
\end{equation}
with $\Pi (\theta)$ being the prior over model parameters $\theta$.
We consider $\log_{10} \mathcal{B}_{\rm vac}^{\rm non-vac} > 1$ to be strong evidence in favor of the non-vacuum hypothesis.
To perform parameter estimation we simulate data and infer upon them using the publicly available codebase \textsc{Bahamas}~\cite{Bahamas,federicopozzoli_2025_16087705}. To compute the evidence, inference is performed using the nested sampling algorithm~\cite{Skilling:2006gxv} as implemented in \texttt{nessai}~\cite{nessai}.

\subsubsection{Dynamical friction}\label{subsec:measurability}
We first consider the effects of dynamical friction by generating injected SGWB data (using methods of~\cref{subsec:eff_DF}) with  $\rho^{\rm inj} \in \{10^{-7}, 10^{-8}, 10^{-9}, 10^{-10}, 10^{-11}\} \mathrm{g \, cm^{-3}}$.
For the RPLP model parameters $\theta_{\rm sBBH}$, we are mainly interested in recovering the vacuum amplitude $A_{\rm vac}$ and the environmental parameter $\alpha$, which in this case is just the disk density $\rho$.
Since $A_{\rm vac}$ and $\gamma$, are strongly correlated, we fix $\gamma = 2/3$ in our inference. 
Likewise, we also set  $\{ A_m, \beta, \kappa \}$ to the values listed in~\cref{tab:env_mapping_params}, due to the strong correlation with $\alpha$.
These parameters are therefore not sampled in our parameter estimation experiments.

We use log-uniform priors $\log_{10} A_{\rm vac} \sim \mathcal{U}(-12, -8)$ to be agnostic regarding the order of magnitude of the vacuum amplitude.
Based on astrophysically motivated values for the disk density~\cite{Barausse:2014pra,Barausse:2014tra}, we take a conservative upper bound on $\rho \lesssim 10^{-6} \mathrm{g} \, \mathrm{cm}^{-3}$, consistent with the perturbative regime of dynamical friction with respect to the Keplerian motion (see Appendix~\ref{app:validity} for more details).
We set log-uniform priors $\log_{10} (\rho/\rho_0) \sim \mathcal{U}(-2, 4)$, with $\rho_0 = 10^{-10} \mathrm{g} \, \mathrm{cm}^{-3}$.
We checked, by further lowering the lower prior bound, that posteriors and evidences are not affected. Therefore we consider our results robust in the limit of $\rho\rightarrow 0$.
For the Galactic foreground parameters, we use the following uniform priors: 
$ \alpha_{\rm vac} \sim \mathcal{U}(0, 2.5)$, $ \log_{10} A_{\rm Gal} \sim \mathcal{U}(-47, -40)$, $ \log_{10} f_{\mathrm{kn}}/{\rm Hz} \sim \mathcal{U}(-4, -2)$, $ \log_{10} f_{1}/{\rm Hz} \sim \mathcal{U}(-4, -2)$, $ \log_{10} f_{2}/{\rm Hz} \sim \mathcal{U}(-4, -2)$.

\begin{figure}[!t]
    \centering
    \includegraphics[width=\columnwidth]{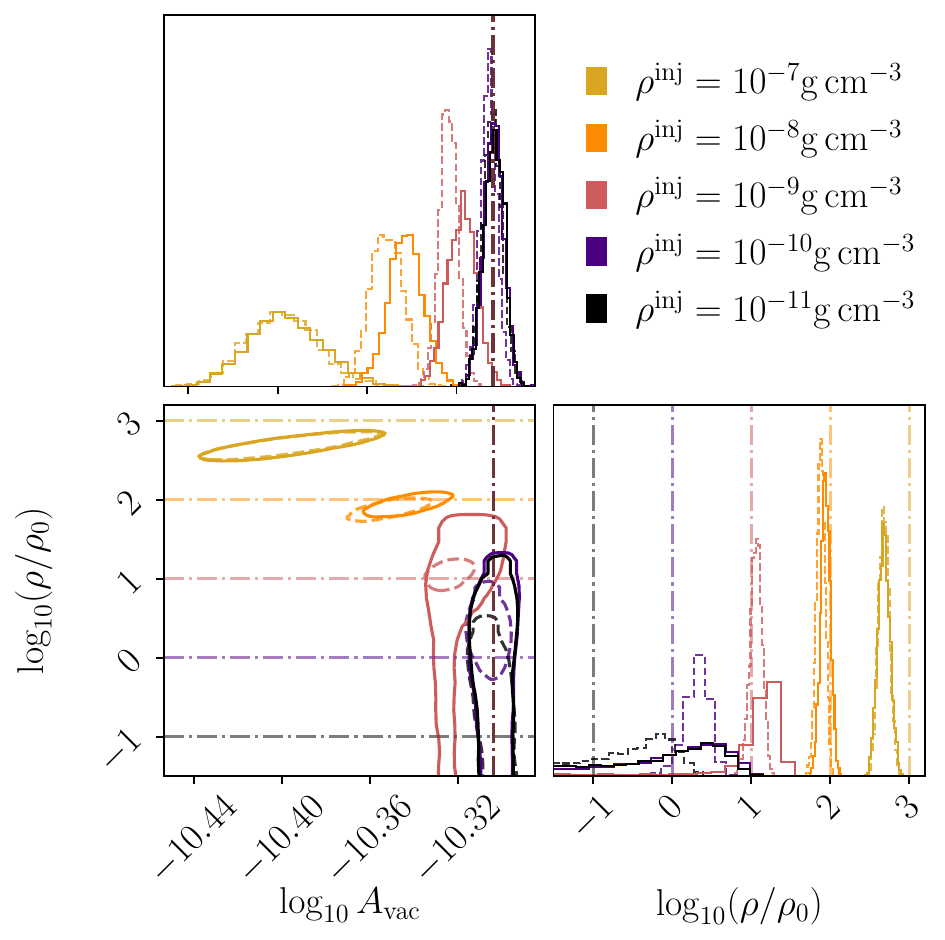}
    \caption{Marginalized posteriors for the dynamical friction model across various matter density regimes. Solid (dashed) contours and histograms correspond to analyses with (without) the inclusion of the Galactic foreground.
    Two-dimensional contours correspond to $90\%$ credible regions.
    Dash-dot represent the true value for $\rho$ and the single asymptotic value for $A_{\rm vac}$ (as listed in~\cref{tab:env_mapping_params}).
    }
    \label{fig:DF_template_recovery}
\end{figure}

In~\cref{fig:DF_template_recovery}, we show the one-dimensional and two-dimensional marginalized posteriors for $A_{\rm vac}$ and $\rho$, inferred with fixed (dashed curves) or free (solid curves) $\theta_{\rm Gal}$.
Overall, we find that the recovered posteriors are well constrained relative to the prior, and the effect of dynamical friction is indeed measurable.

For known Galactic foreground parameters, we observe a trend in the posteriors for increasing $\rho^{\rm inj}$.
When $\rho^{\rm inj} \gtrsim 10^{-9} \mathrm{g} \, \mathrm{cm}^{-3}$, the recovered posteriors of $\rho$ do not overlap, implying that we can better resolve between different orders of magnitude of the disk density.
We quantify this further by computing the statistical uncertainty $\delta \rho/\rho$ using~\cref{eqn:stat_error}.
As listed in~\cref{tab:with_without_galactic}, with increasing $\rho^{\rm inj}$, we find that $\delta \rho$ decreases. 
Specifically, we find that $\delta \rho \lesssim \rho^{\rm inj}$ when $\rho^{\rm inj} \gtrsim 10^{-9} \mathrm{g} \, \mathrm{cm}^{-3}$, with known Galactic foreground parameters.
In all cases, we can infer the vacuum amplitude $A_{\rm vac}$ to $\mathcal{O}(1) \%$ precision.
The fact that we can strongly measure the vacuum amplitude is consistent with the results of~\cite{Babak:2023lro}, which also investigated the detectability of SGWB produced by sBBHs in vacuum.
While the one-dimensional marginalized posterior of $\rho$ shrinks with increasing $\rho^{\rm inj}$, the one-dimensional marginalized posterior of $A_{\rm vac}$ widens instead due to the correlation between the two parameters.
We note that the one-dimensional marginalized posteriors for $\rho$ are biased for increasing values of $\rho^{\rm inj}$. In Appendix~\ref{app:template}, we show that this is due to mismatch between the RPLP model and the true signal caused by the approximate nature of the asymptotic matching used. However, we also show that despite the biases in the marginalized posteriors, the posterior predictive is consistent with the injected signal.

\begin{table}[!t]
\centering
\begin{ruledtabular}
\begin{tabular}{c|cc|cc|cc}
$\rho^{\rm inj}/\rho_0$ 
& \multicolumn{2}{c|}{$\delta \rho/\rho^{\rm inj}$} 
& \multicolumn{2}{c|}{$\delta A_{\rm vac}/A_{\rm vac}^{\rm inj}$} 
& \multicolumn{2}{c}{$\log_{10}\mathcal{B}^{\mathrm{non-vac}}_{\mathrm{vac}}$} \\
\hline
& With & Without & With & Without & With & Without \\
$10^{-1}$ & $37.85$ & $8.218$ & $0.02188$ & $0.01727$ & $1.424$ & $1.010$ \\
$1$      & $4.248$ & $1.4670$ & $0.02075$ & $0.01490$ & $1.146$ & $3.965$ \\
$10$     & $1.118$ & $0.4042$ & $0.02114$ & $0.01430$ & $1.777$ & $8.240$ \\
$10^2$   & $0.3362$ & $0.2033$ & $0.03404$ & $0.02478$ & $1.281$ & $10.16$ \\
$10^3$   & $0.1606$ & $0.1340$ & $0.05334$ & $0.0493$  & $1.427$ & $7.570$ \\
\end{tabular}
\end{ruledtabular}
\caption{Posterior statistical uncertainties (from $90\%$ quantiles) on $\rho$ and $A_{\rm vac}$ together with the Bayes factor between the non-vacuum and vacuum hypotheses. For each injection, results are listed with uncertain (With) or fixed (Without) Galactic foreground parameters. We observe that the precision of $\rho$ improves with increasing $\rho^{\rm inj}$, while that of $A_{\rm vac}$ worsens. Although the Bayes factor decreases significantly when inferring on the Galactic foreground, in all cases the non-vacuum model is strongly preferred.
}
\label{tab:with_without_galactic}
\end{table}

We now discuss the impact of the Galactic foreground parameters on the marginalized posteriors shown in~\cref{fig:DF_template_recovery}.
As expected, the simultaneous inference on Galactic foreground parameters results in wider marginal posteriors. 
For $\rho^{\rm inj} \lesssim 10^{-9} \mathrm{g} \, \mathrm{cm}^{-3}$, however, fixing the Galactic foreground parameters $\theta_{\rm Gal}$ causes an underestimation of $A_{\rm vac}$ and $\rho$ compared to inferring on them.
At higher densities, the influence of Galactic foreground becomes negligible. 
This is related to how the SNR is affected by the Galactic foreground for different densities.

\begin{figure}[b]
    \centering
    \includegraphics[width=\columnwidth]{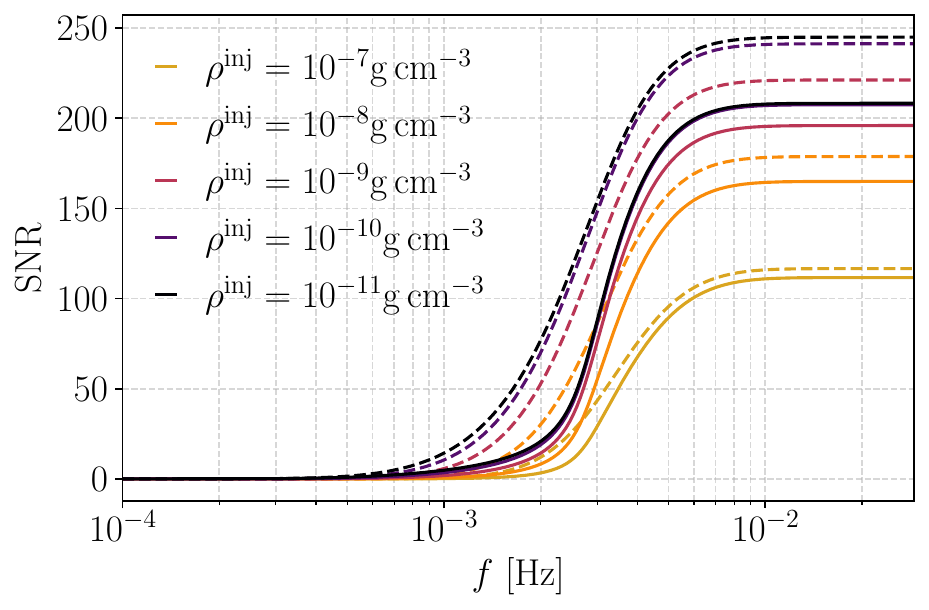}
    \caption{Cumulative SNR of the best-fit recovered SGWB model as a function of frequency. The solid and dashed lines represent the cases with and without the Galactic foreground respectively, while the different colored lines correspond to different $\rho^{\rm inj}$.
    }
    \label{fig:DF_SNR_recovery}
\end{figure}

In~\cref{fig:DF_SNR_recovery}, we show the cumulative SNR with and without including the Galactic foreground parameters, for different values of $\rho^{\rm inj}$.
We compute the SNR of the posterior predictive (see~\cref{app:template} for details).
We find that as $\rho^{\rm inj}$ increases, the total SNR decreases, consistent with the behavior of the SGWB shown in~\cref{fig:DF_gwb}.
Further, including the Galactic foreground lowers the SNR, which along with the additional five Galactic foreground parameters, 
contributes to wider posteriors.
From~\cref{fig:DF_SNR_recovery}, we also observe that with increasing $\rho^{\rm inj}$, the difference in SNR with and without Galactic foreground decreases.
This is because as $\rho^{\rm inj}$ increases, the turning point shifts to frequencies above $\sim 5\times 10^{-3}$Hz, where the Galactic foreground is suppressed.
Consequently, the posteriors of $\rho$ with and without Galactic foreground overlap better as $\rho^{\rm inj}$ is increased, with virtually no difference in the case of $\rho^{\rm inj} \gtrsim 10^{-8} \mathrm{g} \, \mathrm{cm}^{-3}$.
Additionally, the recovered SNR (with and without Galactic foreground) typically accumulates above $50$ only at $f \gtrsim (\mathrm{few}) \times 10^{-3}$Hz.
Thus, the measurements of signal parameters are typically informed by this specific frequency regime.

For each injection, we performed a separate set of runs with a vacuum SGWB model by setting $\alpha = \rho = 0$ in $\Omega_{\rm RPLP}$.
We compute the Bayes factor $\mathcal{B}^{\rm non-vac}_{\rm vac}$ in favor of the non-vacuum model ($\rho \neq 0$), given by~\cref{eq:bayes_factor}, which we list in~\cref{tab:with_without_galactic}.
For all injections, we find that $\mathcal{B}^{\rm non-vac}_{\rm vac}>10$, even when including the Galactic foreground.
Thus, despite the marginalized posteriors for $A_{\rm vac}$ and $\rho$ displaying biases (compared to their vacuum asymptotic and injected values respectively), the RPLP model can detect dynamical friction effects by strongly disfavoring the vacuum model.
We also checked that the posterior predictive from the RPLP model can accurately describe the injected signal within statistical errors, consistent with our Bayes factor results.
We further discuss the biases (as seen in~\cref{fig:DF_template_recovery}) caused by the mismatch between the RPLP model and the true signal in Appendix~\ref{app:template}.

\begin{figure}[t]
    \centering
    \includegraphics[width=\columnwidth]{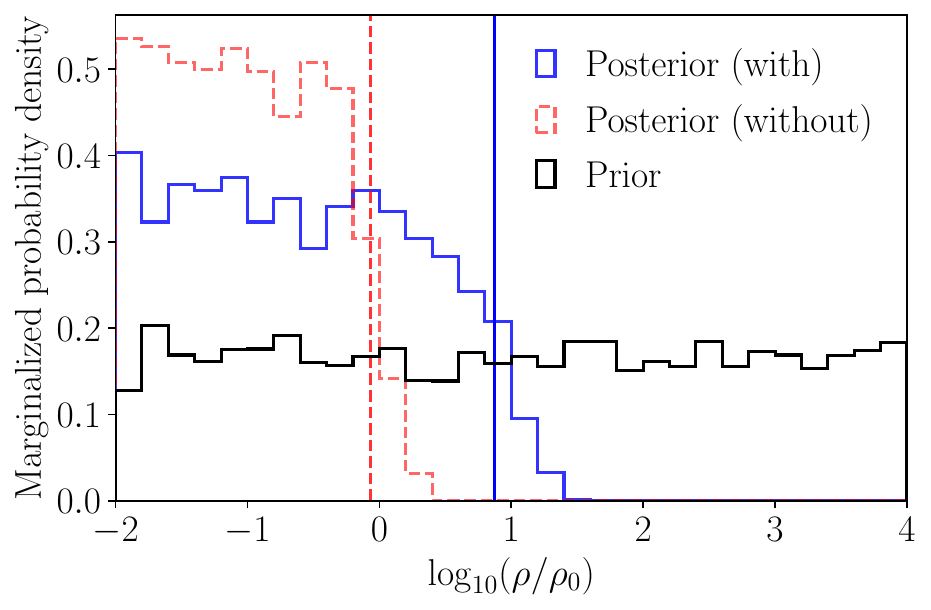}
    \caption{
    Marginalized posterior probability for $\log_{10}(\rho / \rho_0)$ shown for the cases with (blue histogram) and without (red dashed histogram) Galactic foreground parameters included. 
    The vertical lines indicate the one-sided $90\%$ credible intervals and the black histogram indicates the uniform prior on $\log_{10}(\rho / \rho_0)$.
    Observe that the constraint is slightly weaker when including the Galactic parameters.
    }
    \label{fig:constraints_rho}
\end{figure}

If we detect a SGWB consistent with vacuum, our RPLP model can also be used to place upper bounds on the environmental parameters.
In~\cref{fig:constraints_rho}, we show the constraints on $\rho$, with and without including the Galactic foreground parameters in our model.
The constraints, as given by the $90\%$ one-sided credible interval, are informative, because the prior extends  to $\rho = 10^{-6} \mathrm{g \, cm^{-3}}$.
When excluding the Galactic foreground, we find an upper bound of $\rho < 8.51 \times 10^{-11} \mathrm{g}\,\mathrm{cm}^{-3}$.
Owing to the increase in dimensionality when including Galactic foreground, the constraint weakens slightly to $\rho < 7.58 \times 10^{-10} \mathrm{g}\,\mathrm{cm}^{-3}$.

\begin{figure}[t]
    \centering
    \includegraphics[width=\columnwidth]{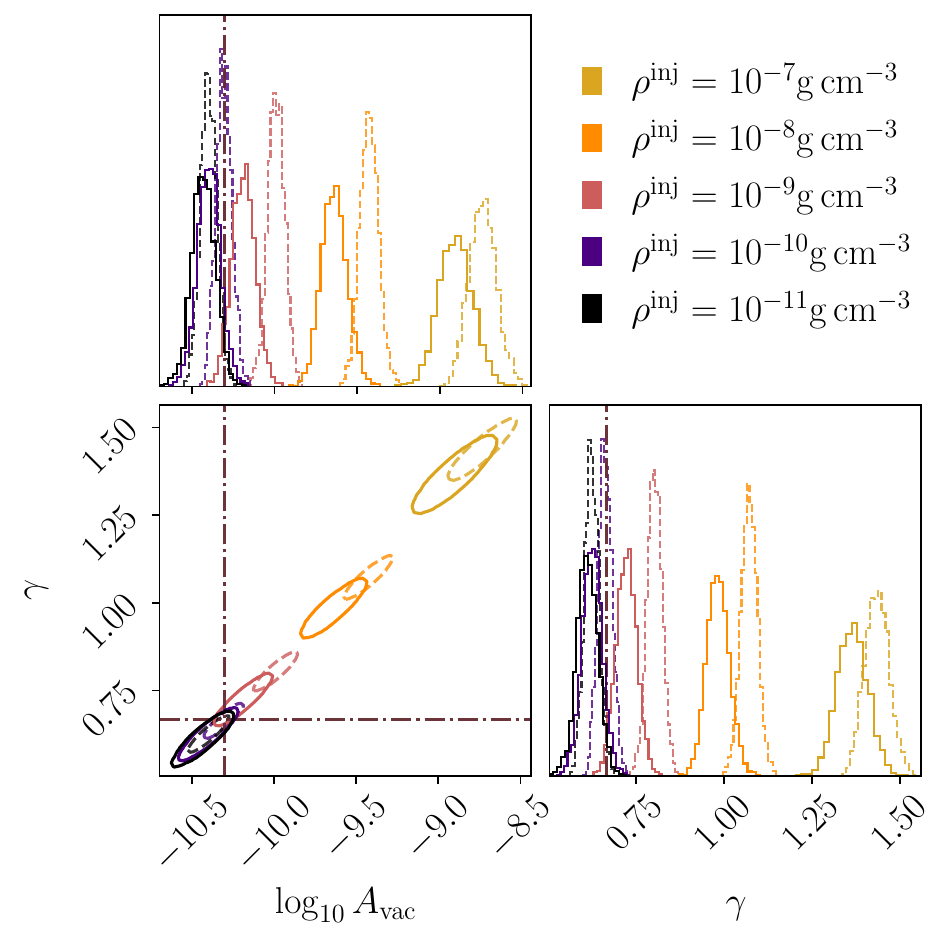}
    \caption{Marginalized posterior probability for the vacuum model across different matter density regimes. Solid and dashed histograms represent the inference results with and without the inclusion of the Galactic foreground, respectively.
    Dash-dotted lines indicate the asymptotic value of $A_{\rm vac}$ (from~\cref{tab:env_mapping_params}) with $\gamma=2/3$.
    Two-dimensional contours correspond to $90\%$ credible regions.
    }
    \label{fig:vac_bias}
\end{figure}
To explore systematic biases on the vacuum SGWB induced by neglecting environmental effects, we inject 
an SGWB containing dynamical friction effects, and analyze it with the vacuum model, i.e.  $\Omega_{\rm RPL}$ with $\alpha = 0$.
In~\cref{fig:vac_bias}, for the injected values of $\rho^{\rm inj} \in \{10^{-7},10^{-8},10^{-9},10^{-10},10^{-11}\}\mathrm{g} \, \mathrm{cm}^{-3}$, we show the recovered two- and one-dimensional marginalized posteriors for the vacuum amplitude $A_{\rm vac}$ and the vacuum spectral index $\gamma$.
When assessing the systematic bias in $A_{\rm vac}$ and $\gamma$, we compare the maximum posterior point to the asymptotic vacuum values (vertical dash-dotted lines) listed in~\cref{tab:env_mapping_params}.
As expected, with increasing $\rho^{\rm inj}$, the systematic biases increase in significance.
In more details, $\gamma$ is biased to values larger than the asymptotic value of $\gamma = 2/3$, because the SGWB containing dynamical friction effects has a steeper slope, with an asymptotic value of 13/3.
Due to the expected positive correlation with $A_{\rm vac}$ (given the functional form of the SGWB), we find that $A_{\rm vac}$ is biased to larger values.
When inferring simultaneously on Galactic foreground parameters, due to the reduced constraining power on dynamical friction show in~\cref{fig:DF_template_recovery}, the biases on $A_{\rm vac}$ and $\gamma$ are milder, as expected.

\subsubsection{Gas accretion}
\begin{figure}[!ht]
    \centering
    \includegraphics[width=\columnwidth]{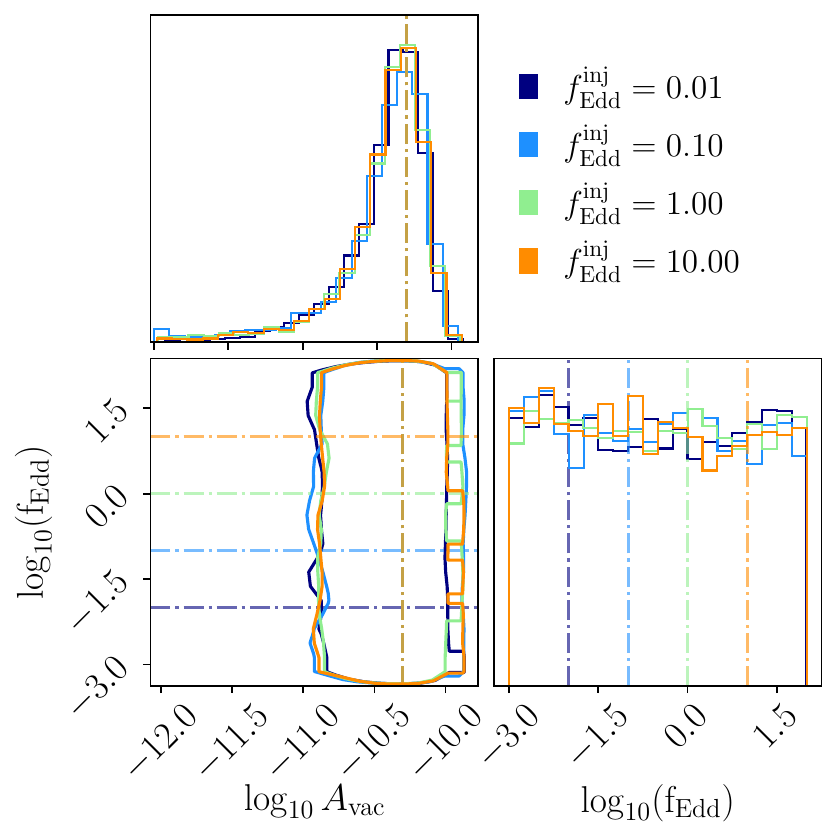}
    \caption{Marginalized posterior probability for the gas accretion model across different Eddington ratios. 
    Dash-dot lines represent the true value of $f_{\rm Edd}$ and the asymptotic value of $A_{\rm vac}$ (as listed in Table~\ref{tab:env_mapping_params}).
    Two-dimensional contours correspond to $90\%$ credible regions.
    Observe that the posteriors of $f_{\rm Edd}$ are identical to the prior, implying gas accretion effects cannot be inferred.
    }
    \label{fig:edd_recovery}
\end{figure}
 
To characterize gas accretion measurability, we inject SGWB data 
(using models in~\cref{subsec:eff_acc})
with $f_{\rm Edd}^{\rm inj} \in \{0.01, 0.10, 1.00, 10\}$.
We use a log-uniform prior given by $\log_{10} f_{\rm Edd} \sim \mathcal{U}(-3, 2)$, where the upper bound reflects astrophysical expectations of how large the Eddington ratio can be. We have checked that our results are robust to the exact choice of the lower bound, which we cannot set exactly to zero due to the log-uniform prior.
We carry out the Bayesian inference just like we did for dynamical friction by fixing $\{\gamma,A_m, \beta, \kappa \}$.

In~\cref{fig:edd_recovery}, we show the two-dimensional and one-dimensional marginalized posteriors of $A_{\rm vac}$ and $f_{\rm Edd}$. 
We accurately measure the vacuum amplitude across all injections, thus finding that marginalizing over gas accretion does not impact the measurability of the vacuum stellar SGWB.
However, we do not obtain informative posteriors of Eddington ratios for any of the injected values, even with $f^{\rm inj}_{\rm Edd} = 10$ marginally recovering the prior.
The lack of constraining power on $f_{\rm Edd}$ is due to the turning point of the SGWB occurring at frequencies lower that the LISA sensitive band, as we anticipated qualitatively with BPLSs in~\cref{subsec:eff_acc}.
Given our findings, we do not perform any additional Bayesian analysis for gas accretion effects.

\subsection{Measurability of dynamical friction effects from a sub-population of sBBHs}\label{subsec:env_fraction}
\begin{figure}[t]
    \centering
    \includegraphics[width=\columnwidth]{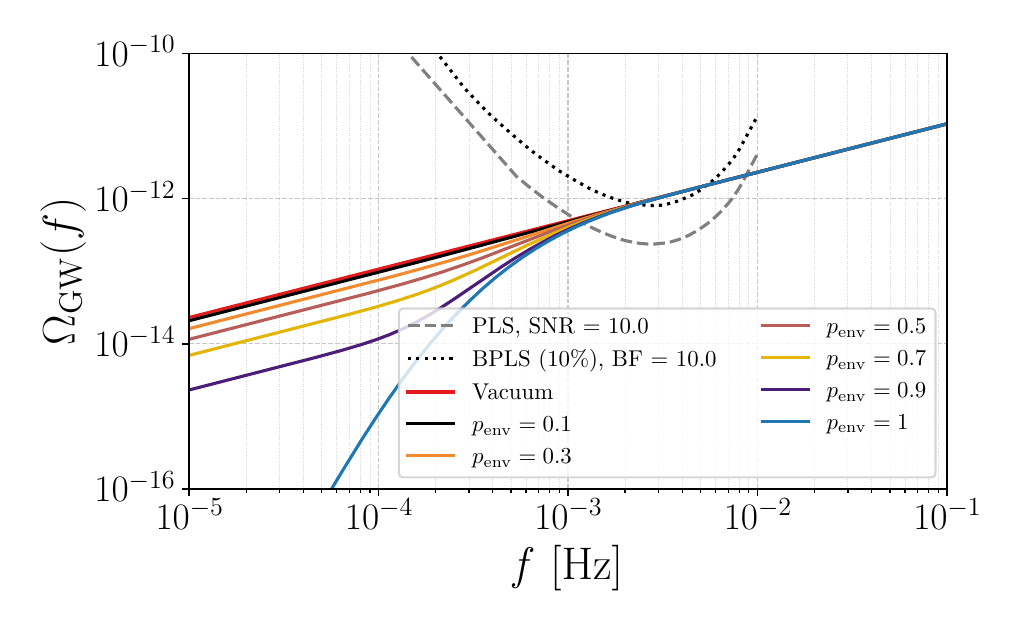}
    \caption{
SGWB spectra $\Omega_{\mathrm{GW}}(f)$ for various environmental fractions $p_{\mathrm{env}}$ in the case of dynamical friction with $\rho^{\rm inj} = 10^{-10} \mathrm{g \, cm^{-3}}$.
The gray dashed and black dotted curves indicate the power-law sensitivity (PLS) and Bayesian power-law sensitivity (BPLS) of LISA, under the same assumptions in~\cref{fig:Acc_gwb,fig:DF_gwb}.
Here, BF denotes the Bayes factor of BPLS.
}
    \label{fig:DF_fraction}
\end{figure}

The impact of the SGWB due to environmental effects will depend on the distribution of the environmental parameters ($\rho$ and $f_{\rm Edd}$) across the population. 
In a realistic scenario, only a sub-population of sBBHs formed in gaseous medium, where the environmental effects are significant.
As a proof-of-concept, we phenomenologically construct a mixture model to account for environmental effects arising from a sub-population.
We introduce the fraction parameter $p_{\mathrm{env}}$, which characterizes the fractional contribution of the SGWB arising from binaries affected by environmental effects.
The total SGWB is then given by
\begin{equation}
\label{eqn:omega_frac}
\Omega_{\mathrm{Frac}} = p_{\mathrm{env}} \Omega_{\mathrm{env}}+(1-p_{\mathrm{env}})\Omega_{\mathrm{vacuum}},
\end{equation}
where $\Omega_{\mathrm{env}}$ and $\Omega_{\mathrm{vacuum}}$ correspond to the SGWB spectra of the two mixture components: one in which all binaries are affected by environmental effects, and the other in which all binaries evolve in vacuum, respectively.
The physical interpretation for our model need not be unique, however it can represent, e.g., a sBBH population whose binaries above (or below) a certain mass are affected by environmental effects.

In~\cref{fig:DF_fraction}, for a given injected density of $\rho^{\rm inj} = 10^{-10}\mathrm{g} \, \mathrm{cm}^{-3}$, we illustrate how changing the injected fraction $p_{\rm env}^{\rm inj}$ leads to  changes in the SGWB.
The case of $p_{\rm env} = 0$ is identical to the vacuum SGWB, while  $p_{\rm env} = 1$ is identical to having all binaries affected by environmental effects (as shown in~\cref{fig:DF_gwb}).
Notably, for $0<p_{\rm env} < 1$ the total SGWB exhibits a spectral shape resulting from the $p_{\rm env}$-weighted superposition of two SGWBs, dominated by vacum component at low frequencies.
If future LVK observations provide tighter constraints on the binary population, the difference in amplitude of the SGWB at high and low frequencies might signal the presence of a sub-population containing environmental effects.

With only a sub-population affected by environmental effects, we assess how the measurability of $\rho$ changes with $p_{\rm env}$.
To do so, we first generate different injections of $\Omega_{\rm Frac}$ using~\cref{eqn:omega_frac} for $p^{\rm inj}_{\rm env} \in \{ 0.1, 0.3, 0.5, 0.7, 0.9 \}$ and a fixed density $\rho^{\rm inj} = 10^{-10} \mathrm{g}\, \mathrm{cm}^{-3}$.
Specifically, in~\cref{eqn:omega_frac}, we evaluate $\Omega_{\rm env}$ and $\Omega_{\rm vac}$ as integrals over the population, which we discussed in Sec.~\ref{subsec:eff_DF} and Sec.~\ref{sec:SGWB_vac} respectively.
We then perform parameter estimation using $\Omega_{\rm Frac}$, but with $\Omega_{\rm env}$ and $\Omega_{\rm vac}$ evaluated using the RPLP model.
Note that the RPLP model describes $\Omega_{\rm vac}$ with $\alpha = 0$.
As in Sec.~\ref{subsec:measurability}, we perform the analysis with and without the Galactic foreground.

\begin{figure*}[t]
    \centering
    \includegraphics[width=0.7\linewidth]{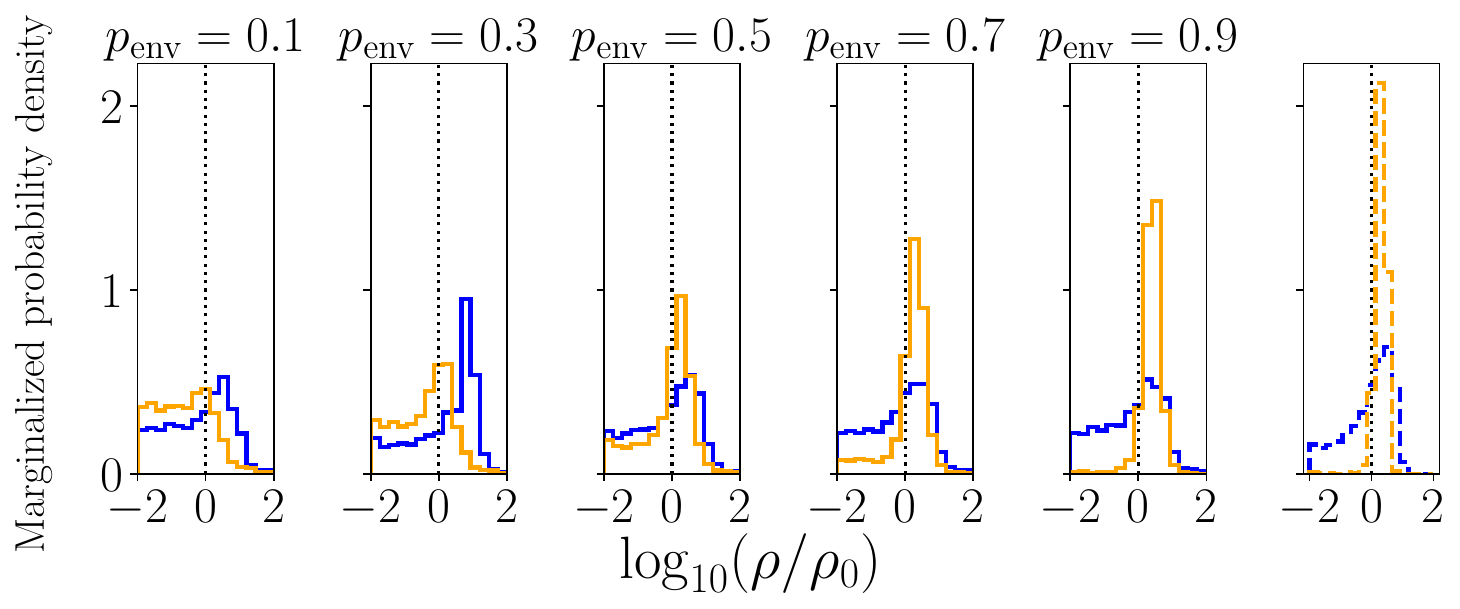}
    \caption{
    {Marginalized posterior distribution of $\log_{10}(\rho / \rho_0)$ as a function of the environmental fraction parameter $p_{\rm env}$, with an injected density $\rho^{\rm inj} = 10^{-10} \rm{g \, cm^{-3}}$ (horizontal dashed line). The first five panels show the recovered posteriors using the mixture model for different injected values of $p_{\rm env}$. The last panel is for reference and shows the recovered posteriors using the RPLP model when injecting with $p_{\rm env}=1$ (as shown in~\cref{fig:DF_template_recovery}).}
    } 
    \label{fig:posterior_mixed_fraction}
\end{figure*}

In~\cref{fig:posterior_mixed_fraction}, the blue (orange) histograms correspond to the one-dimensional marginalized posteriors of $\rho$ with (without) including the Galactic foreground parameters.
The first five panels of~\cref{fig:posterior_mixed_fraction} correspond to the cases with different injected fraction values. 
Meanwhile, for reference, the last panel of~\cref{fig:posterior_mixed_fraction} shows the case where the injected data contains dynamical friction effects for all binaries with $p_{\rm env} = 1$, which we had shown earlier in~\cref{fig:DF_template_recovery}. 
We find that (for this injected density) the impact of the injected fraction on the marginalized posterior of $\rho$ is not significant  when including the Galactic foreground.
This is due to the fact that the Galactic foreground is not suppressed in the regime where the signal undergoes a turning point as a result of dynamical friction. 
When excluding the Galactic foreground, we can better see the impact of the sub-population on the marginalized posterior of $\rho$.

For each injection, we also separately infer with the vacuum model and compute the log Bayes factor $\log_{10} \mathcal{B}^{\rm frac}_{\rm vac}$ between the mixture and vacuum models. 
We find that for all values of the injected fraction, there is substantial preference for the mixture model over the vacuum model with Bayes factors in the range $\log_{10} \mathcal{B}^{\rm Frac}_{\rm vac} \sim 0.79$---$0.95$ depending on the fraction.
When excluding the Galactic foreground, we instead have $\log_{10} \mathcal{B}^{\rm Frac}_{\rm vac} \sim 1.27$---$3.52$ depending on the fraction, indicating strong preference for the mixture model. 
In other words, the evidence in favor of the mixture model (over vacuum) is weaker when including the Galactic foreground, as we had also seen in previous analysis with $p_{\rm env}=1$.
We have thus shown that even with only sub-populations undergoing dynamical friction effects, and with including the Galactic foreground, we can reject the vacuum model with substantial evidence.%

\section{Agnostic tests of additional energy loss channels}\label{subsec:agnostic}
In general, when there is energy loss from binaries in addition to GW emission, the SGWB will decrease below the vacuum GR prediction given by~\cref{eqn:dEdf_vac}.
We use our phenomenological model to probe such additional energy losses agnostically.
Specifically, with our RPL model (see~\cref{eqn:RPL_template}), one can constrain the dimensional combination $\tilde{\alpha} \equiv A_m \alpha$ for different choices of $\{\beta,\kappa \}$.
In the intermediate-frequency regime, since the RPL model typically overestimates the signal relative to the RPLP model, the constraints obtained with the RPL model would thus be conservative.
In the context of environmental effects, equipped with a mapping of the parameters and a population, one can place upper bounds on specific environmental parameters.
However, the RPL model can also be used to test several modified gravity theories with an appropriate mapping.
This is similar to~\cite{Maselli:2016ekw}, where the authors used the parameterized post-Einsteinian (ppE) framework to place constraints on several modified theories of gravity.
The advantage of our models over the ppE model of~\cite{Maselli:2016ekw} is that we do not assume that the additional energy flux is small relative to the Newtonian GW flux.
When the additional energy dissipation is small, results from our model will agree with those of~\cite{Maselli:2016ekw}.
To demonstrate  our approach, we consider the case of a time-dependent Newton gravitational constant $G(t)$ (see~\cite{Yunes:2009bv,Tahura:2018zuq}) and obtain an order of magnitude upper bound on how the SGWB from sBBHs can place independent constraints on $|\dot{G}/G|$. 

The effect of a linearly time-varying $G(t)$ behaves identically to gas accretion, and thus is also a -4PN effect relative to GW radiation reaction. 
Comparing the expressions for $\dot{f}$ in the two cases, the effects can be mapped upon the substitution $f_{\rm Edd}(5+3\xi)/\tau \rightarrow 2\dot{G}_0/G_0$, where $G$ and $\dot{G}$ are evaluated at a reference time $t=t_0$.
We note here that the original result for $\dot{f}$ in~\cite{Yunes:2009bv} is incorrect, because the authors apply energy balance (which is not warranted), while the correct expression is easily obtained using angular momentum balance, as done in~\cite{Tahura:2018zuq} (see also~\cite{Caputo:2020irr} for the case of gas accretion). 
While the incorrect application of energy balance and the result from~\cite{Yunes:2009bv} have been used to place constraints on $|\dot{G}_0/G_0|$ (cf. e.g.~\cite{Yunes:2016jcc,Chamberlain:2017fjl,Barbieri:2022zge,Cannizzaro:2023mgc}), the constraints' order of magnitude is not spoiled (see~\cite{Tahura:2018zuq} and Appendix~\ref{app:edd} for further discussion).

A deviation from the vacuum GR prediction for the SGWB is constrainable if the turning point of the spectrum occurs in the detector sensitivity band. 
In Sec.~\ref{subsec:measurability}, we found that using an astrophysical upper bound on the Eddington ratio with $f_{\rm Edd}\leq 100$, we could not constrain the effect of accretion, due to the turning point being at frequencies lower than the LISA sensitivity band.
We now estimate the upper bound on $|\dot{G}_0/G_0|$ by requiring that the turning point of the signal be at $f \sim 10^{-3}$Hz, which is where LISA  is optimally sensitive. 
We find that when $|\dot{G}_0/G_0| > 10^{-4} \mathrm{yr}^{-1}$, the turning point would be above $10^{-3}$Hz, resulting in a $\gtrsim 10 \%$ change in the vacuum GR signal, which can  be tested with LISA.
This order of magnitude constraint of $|\dot{G}_0/G_0| \lesssim 10^{-4} \mathrm{yr}^{-1}$ is nearly 10 orders of magnitude weaker than that from lunar ranging measurements~\cite{Genova2018} (see~\cite{Will:2014kxa} for a review).
However, as an independent constraint, it is comparable to what can be obtained by the loudest (and best known) verification Galactic binary in the LISA band~\cite{Barbieri:2022zge}, and slightly better than what neutron star mergers can constrain with future third-generation  detectors~\cite{Chamberlain:2017fjl}.
Our estimate is also consistent with~\cite{Cannizzaro:2023mgc}, when appropriately rescaling their results to mHz frequencies and stellar mass binaries.
Thus, not only can the detection of SGWB of sBBHs in the LISA band probes environmental effects, but it also allows for independent and agnostic tests of GR, both of which can be accomplished using the phenomenological models developed in this work.

\section{Conclusion} \label{sec:conclusion}
The SGWB from sBBHs, which can be detected by LISA, offers a probe of the astrophysical environment in which these binaries form. 
This is enabled by a suppression of the SGWB due to additional energy-loss channels induced by the surrounding environment.
In this work, we investigated the detectability (with LISA) of dynamical friction and gas accretion on the SGWB of sBBHs.
Assuming that all sBBHs undergo either dynamical friction or gas accretion, our major results are that (i) for typical disk densities $\rho \sim 10^{-10}-10^{-9} \mathrm{g \, cm^{-3}}$, dynamical friction effects are well measured and the vacuum model is confidently disfavored even when including the Galactic foreground in the inference, (ii) gas accretion cannot inferred or constrained even for an Eddington ratio of $f_{\rm Edd} = 10$ even excluding the Galactic foreground from the inference.
As a consequence of (i), we also found that there are significant systematic biases in the inference of the vacuum amplitude and spectral index, highlighting the need to model environmental effects in the SGWB.

Further, we investigated the impact of a sBBH sub-population undergoing dynamical friction effects. 
As a proof-of-concept, using a mixture model, we showed that even when only a sub-population of sBBHs undergoes dynamical friction with moderate values of the disk density, LISA can substantially reject the vacuum model, while still including the Galactic foreground parameters.
We constructed phenomenological parametric models that can capture environmental effects, and more generally, any additional energy loss channel.
Thus our models can also be used to test GR, provided the modifications to GR are dominated by dissipative effects. 
As a proof-of-concept, we obtained an order-of-magnitude constraint on the time variation of Newton's constant: $|\dot{G}/G| \lesssim 10^{-4} \mathrm{yr}^{-1}$, competitive with the projected constraint from the loudest LISA verification binary~\cite{Barbieri:2022zge}.

As a first pass, we assumed that all non-vacuum sBBHs have the same environmental effect with identical values for the disk density and Eddington ratio.
In reality, one needs to also model the population distribution of the environmental parameters, which will affect the resulting SGWB.
We also neglected eccentricity in the modeling of the sBBH orbits.
As shown in~\cite{Chen:2016zyo}, the SGWB is suppressed at low frequencies due to eccentricity.
Thus, the effects of a vacuum eccentric sBBH population could be degenerate with that of a non-vacuum quasi-circular sBBH population, which we will explore as future work.
In addition, environmental effects themselves can affect eccentricity evolution~\cite{Barausse:2007dy,Gair:2010iv,Bonetti:2017dan,Cardoso:2020iji,Naoz:2012bx,Will:2013cza,Chandramouli:2021kts}. 
Specifically, in the context of our work, it is important to characterize how disk-induced dynamical friction and gas accretion can influence the eccentricity distribution of a non-vacuum sBBH population.

In our modeling of the stochastic signal in the LISA data, we neglected the contribution from the extragalactic white dwarf population~\cite{farmer_phinney,staelens_nelemans}, which can potentially affect the detection of the SGWB from the sBBH population.
Optimistically, an LVK detection of the sBBH stochastic signal can help detect the same in the LISA band, even in the presence of an extragalactic white dwarf SGWB, which we shall explore as future work.
Finally, the phenomenological parametric models we developed can be improved with a better modeling of the asymptotic and intermediate regimes, so as to provide more accurate inference of the environmental parameters and the posterior predictive of the stochastic signal.

\begin{acknowledgements}

The authors wish to thank Cole Miller, Laura Sberna, Alberto Sesana, and Rodrigo Vicente for fruitful discussions.
R.C is supported by China Scholarship Council (No.202406340069), the Natural Science Foundation of China (No.12233011).
R.B. acknowledges support from the ICSC National Research Center funded by NextGenerationEU, and the Italian Space Agency grant Phase B2/C activity for LISA mission, Agreement n.2024-NAZ-0102/PER. 
We acknowledge support from the European Union’s H2020 ERC Consolidator Grant ``GRavity from Astrophysical to Microscopic Scales'' (Grant No. GRAMS-815673, to E.B. and R.S.C.), the European Union’s Horizon ERC Synergy Grant ``Making Sense of the Unexpected in the Gravitational-Wave Sky'' (Grant No. GWSky-101167314, to E.B.), the PRIN 2022 grant ``GUVIRP - Gravity tests in the UltraViolet and InfraRed with Pulsar timing'' (to E.B. and R.S.C.), and the EU Horizon 2020 Research and Innovation Programme under the Marie Sklodowska-Curie Grant Agreement No. 101007855 (to E.B.).
This work has been supported by the Agenzia Spaziale Italiana (ASI), Project n. 2024-36-HH.0, ``Attività per la fase B2/C della missione LISA''.  
The data underlying this article will be shared on reasonable request to the corresponding authors.

\end{acknowledgements}
\appendix

\section{Merger rate and population model}\label{app:pop_model}
The local merger rate of binary black holes near redshift $z = 0$ is relatively well constrained~\cite{KAGRA:2021duu}. However, its evolution at higher redshifts remains highly uncertain~\cite{KAGRA:2021kbb, KAGRA:2021duu}.
We assume that the merger rate density of gravitational wave sources, $R_{\mathrm{GW}}(z)$,  as being proportional to the cosmic star formation rate (SFR) $R_{\mathrm{SFR}}$~\cite{Madau:2014bja}, expressed as
\begin{equation}
R_{\mathrm{SFR}}(z)=\frac{R_0}{\mathcal{C}} \frac{(1+z)^{\lambda_{1}}}{1+\left(\frac{1+z}{1+z_p}\right)^{\lambda_{1}+\lambda_{2}}}.
\end{equation}
Here, $R_{0}$ denotes the local rate of binary systems at redshift $z = 0$, while $\mathcal{C}$ is a normalization constant ensuring $R_{\mathrm{GW}}(0) = R_{0}$. 

As sBBH formation is expected to be more efficient in low-metallicity environments, we weigh the SFR by the fraction of stars formed with metallicity below a critical threshold,
$F\left(Z<Z_{\mathrm{thresh}}, z\right)$.
This fraction follows the fitting formula of Ref.~\cite{Langer:2005hu}, and we adopt a more stringent
cutoﬀ $Z_{\mathrm{thresh}}=0.1 Z_{\odot}$~\cite{Chruslinska:2018hrb,Mapelli:2019bnp}.
Moreover, black holes are expected to undergo a range of evolutionary time delays between formation and their eventual binary mergers.
We assume these time delays follow a log-uniform distribution, $p\left(t_d\right) \sim t_d^{-1}$, within the range $10~\mathrm{Myr} \leq$ $t_d \leq 13.5~\mathrm{Gyr}$~\cite{Callister:2023tws}.
The unnormalized merger rate is then obtained by convolving this distribution with the metallicity-weighted star formation rate: 
\begin{align}
\tilde{R}_{\mathrm{GW}}(z)=\int d t_d R_{\mathrm{SFR}}\left(z_{f}\right)F\left(Z<Z_{\mathrm{thresh}}, z_f\right) p\left(t_d\right),
\end{align}
where $z_f \equiv z_f(z,t_d)$ denotes the redshift corresponding to the formation time of the binary black hole.
Finally, we normalize the merger rate as
\begin{align}
R_{\mathrm{GW}}(z)=R_0 \frac{\tilde{R}_{\mathrm{GW}}(z)}{\tilde{R}_{\mathrm{GW}}(0)}.
\end{align}

Recent observations indicate that the sBBH population exhibits low effective spins~\cite{LIGOScientific:2020kqk,Miller:2020zox}.
Consequently, when computing the SGWB, we assume that sBBHs have negligible spin and focus solely on the mass distribution, which is described by the Power Law + Peak mode~\cite{Talbot:2018cva}.
Under this assumption, the source parameters $\boldsymbol{\phi}$ in~\cref{eqn:Omega_GW_calculation} contain the primary mass $m_1$ and mass ratio $q$, and the distribution $p(\boldsymbol{\phi})$ reduces to
joint probability density function $p(m_1,q|~{\mathrm \Lambda_{m}})$, where ${\mathrm \Lambda_{m}}=\left\{\alpha_{\rm pop}, \beta_{\rm pop}, m_{\min}, m_{\max}, \delta_{\rm m}, \lambda_{\rm peak},\mu_{\rm m},\sigma_{\rm m}\right\}$ represents the set of hyperparameters governing the mass distribution.

\begin{figure}[!h]
\centering
\includegraphics[width=\columnwidth]{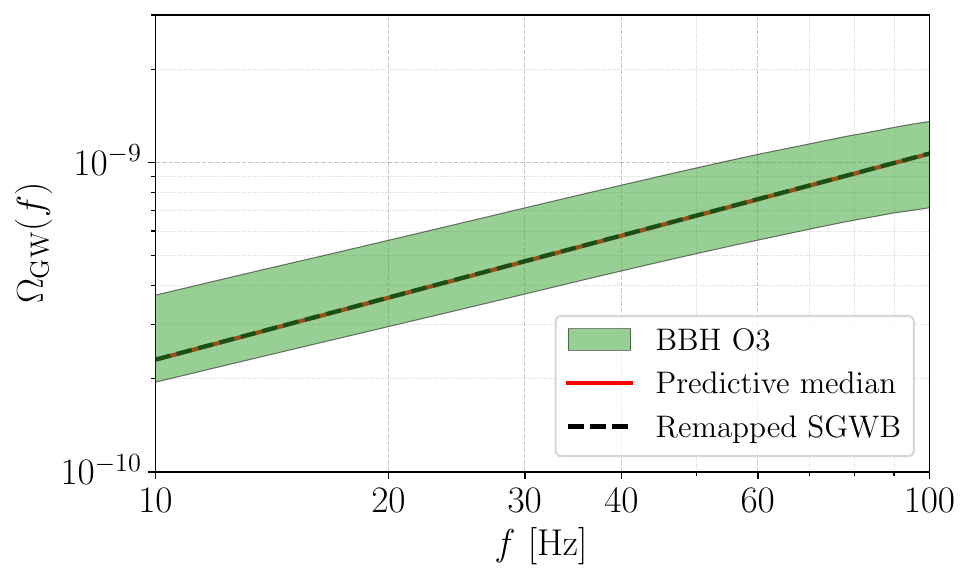}
\caption{
The predictive median SGWB spectrum  generated by our adopted model of sBBH mergers (red solid) and the remapped spectrum based on fiducial parameter values (black dotted), compared with the predicted 90\% credible spectrum band (green) from the O3 observational data~\cite{KAGRA:2021duu} in the frequency band of 10-100 Hz.
}
\label{fig:consistency with O3}
\end{figure}

We numerically evaluate~\cref{eqn:Omega_GW_calculation} via Monte Carlo integration, using the corresponding posterior distributions of merger rate and mass distribution models inferred from~\cite{KAGRA:2021kbb,KAGRA:2021duu} and generate posterior predictions for the SGWB. 
We obtain the predictive posterior median by evaluating, at each frequency, the median of the SGWB realizations from the posterior samples.
By computing the $\ell^2$-norm between each sampled SGWB spectrum and the median curve, we select the sample that minimizes this distance and the corresponding parameters are adopted as fiducial values in this work.
The merger rate parameters are taken as $R_0 = 21~\mathrm{Gpc}^{-3} \mathrm{yr}^{-1}$, $\lambda_1 = 1.5$, $\lambda_2 = 3.7$, and $z_p = 3.0$.
The hyperparameters of mass distribution are $\alpha_{\rm pop} = 3.6$, $\beta_{\rm pop} = 3.4$, $m_{\min} = 4.6 M_\odot$, $m_{\max} = 84 M_\odot$, $\delta_{\rm m} = 4.5M_\odot$, $\lambda_{\rm peak} = 0.018$, $\mu_{\rm m} = 29,M_\odot$, and $\sigma_{\rm m} = 8.3,M_\odot$.

In~\cref{fig:consistency with O3}, we demonstrate the consistency between the median predicted SGWB curve generated by our adopted model and the forecast based on the O3 observational data~\cite{KAGRA:2021duu}.
The remapped spectrum corresponding to the fiducial parameter set closely tracks the predictive median, and both lie well within the $90\%$ credible interval of the O3 inference.
Further, we observe that the spectral indices match well at low frequencies corresponding to the inspiral regime. 
However, at higher frequencies, higher order PN effects become relevant as the spectrum transitions toward merger and ringdown regimes. 
Since we use a leading PN approximation to compute the energy sepctrum given by~\cref{eqn:dEdf_vac}, we find that the spectral index is overestimated relative to the O3 inference that uses the higher PN terms. 
In the lower-frequency bands relevant to LISA, which is our primary focus, the energy spectrum of sBBHs is well described by the leading PN expression of~\cref{eqn:dEdf_vac}, and thus we are justified in neglecting higher PN terms for this work.

\section{Validity of dynamical friction and gas accretion modeling} \label{app:validity}

\subsection{Validity of dynamical friction modeling}

Requiring the dynamical friction force to be perturbative is equivalent to requiring that the dynamical friction timescale $\tau_{\rm DF}$ is much longer than the Keplerian orbital timescale $\tau_{\rm orb}$. 
In a multiple-scale-analysis treatment~\cite{benderorszag}, we essentially consider dynamical friction to leading order in $\epsilon_{\rm DF} \sim \tau_{\rm orb}/ \tau_{\rm DF}$.
We can see that  $\epsilon_{\rm DF}$ scales with $v^{-6}$, making it a $-3$PN relative contribution to the Keplerian equations of motion.
However, being a dissipative effect, the dynamical friction force scales as $v^{-11}$ relative to the GW radiation reaction force, making it a $-5.5$PN contribution at the level of the GW flux.
The important point here is that the perturbative $\epsilon_{\rm DF} \ll 1$ regime does not necessarily imply that dynamical friction is perturbative relative to GW radiation reaction, as is typically assumed when computing the GW phase~\cite{Toubiana:2020drf}. 

For a typical stellar binary with comparable masses, in the very early inspiral regime of $f_s \sim 10^{-5}$Hz, dynamical friction will become non-perturbative when $\epsilon_{\rm DF} \sim 1$, resulting in
\begin{align}
    \rho > (10^{-5} \mathrm{g} \, \mathrm{cm}^{-3}) \left ( \dfrac{f_s}{10^{-5}\mathrm{Hz}} \right)^2 \left( \dfrac{\ln[1/(10^{-5}\mathrm{Hz})]}{\ln(1/f_s)} \right),
\end{align}
where we have also neglected the dependence on $f_{A,*}$ (for the same reasons as in the estimate of $f_{\rm turn,DF}$).
Typically, we expect perturbation theory to become inaccurate at an even smaller density, when $\epsilon_{\rm Df}\lesssim 1/10$, which gives a conservative upper bound:  
\begin{align}
\rho_{\rm upper, modeling}<10^{-6} \mathrm{g} \, \mathrm{cm}^{-3}.  \label{eq:rho_modeling} 
\end{align}
We find that the agnostic upper bound from~\cref{eq:rho_modeling} is consistent with typical densities at the migration trap of thin Keplerian accretion disks around supermassive black holes (SMBHs) as discussed in~\cite{Barausse:2014tra}.
Thus, for densities with $\rho \ll \rho_{\rm upper, modeling}$, using linear perturbation theory is sufficient. 
In our analysis, we consider values of $\rho \lesssim 10^{-7} \mathrm{g} \, \mathrm{cm}^{-3}$, which corresponds to $\epsilon_{\rm DF} \lesssim 10^{-2}$ that shows that we are well within the perturbative regime. 

Due to the presence of the log-term in the dynamical friction force model of~\cite{Kim:2007zb}, it naively appears that there is a typical frequency $f_A^* \sim 50 v_s/11\pi m_A$ above which the force aids the motion and leads to anti-chirping contribution to the frequency evolution.
At those frequencies, the corresponding Mach number is well outside the values considered by~\cite{Kim:2007zb}, and thus their model is not reliable.
In fact, at those high frequencies, for accurate modeling of the dynamical friction effects, one has to also include the force due to the wake created by body B on body A, which becomes increasingly relevant as the two bodies inspiral closer to each other.

Furthermore, relativistic effects such as 1PN corrections to the dynamical friction evolution (formally scaling as $\mathcal{O}(v^2 \epsilon_{\rm DF})$ relative to the Keplerian dynamics) will also become increasingly relevant as the binary inspirals.
These relativistic corrections are relevant for predicting the long-time evolution of a single binary, e.g.  the evolution of eccentricity as shown in the toy model of~\cite{Gair:2010iv}.
However, for predicting the SGWB, at the frequencies where~\cref{eqn:DF_force} breaks down or mutual wake effects or relativistic couplings become relevant, GW emission will completely dominate the effect of dynamical friction, simply due to the relative frequency scaling of $f^{-11/3}$. 
Essentially, at higher frequencies the GW energy spectrum will rapidly asymptote to the vacuum SGWB, and higher order modeling of dynamical friction effects (that only matter at high frequencies) will not affect the prediction of the SGWB.
The main impact of dynamical friction is at lower frequencies, where~\cref{eqn:DF_force} is valid, and mutual wake effects and relativistic couplings can be safely neglected.

\subsection{Validity of gas accretion modeling}\label{app:edd}
The effects of accretion are perturbative relative to the Keplerian motion when $\epsilon_{\rm Acc} \sim T_{\rm orb} \dot{f}_s/f_s \sim T_{\rm orb} \dot{m}/m \ll 1$.
For typical values of the drag coefficient $\xi \sim \mathcal{O}(1)$, we obtain an upper bound on $f_{\rm Edd}$ by requiring $\epsilon_{\rm Acc} <1/10$, which results in
\begin{align}
\begin{aligned}
        f_{\rm Edd} &\lesssim \left(1.4 \times 10^7\right) \left( \dfrac{f_s}{10^{-5} \, \mathrm{Hz}} \right) \left( \dfrac{\tau}{4.5 \times 10^{7} \, \mathrm{yr}} \right) \left( \dfrac{8}{5+3\xi} \right).
\end{aligned}
\end{align}  
Thus for astrophysical values of $f_{\rm Edd} \sim 0.01$--$100$, the effects of accretion are well within the perturbative regime.

We now present a pedagogical discussion of some theoretical issues concerning the derivation of the back-reaction under gas accretion and GW emission.
The main issue stems from the fact that the masses are time-dependent which causes a subtlety in applying the flux balance laws.
We introduce $\epsilon_{\rm RR} \sim \tau_{\rm orb} / \tau_{\rm RR}$ with $\tau_{\rm RR}$ being the GW radiation reaction timescale.

Naively, one might write down the energy balance law as
\begin{equation}
\dot{E}_b =  - \epsilon_{\rm RR} \dot{E}_{\rm GW} - \epsilon_{\rm Acc} \dot{E}_{\rm Acc},\label{eqn:e-balance-wrong}   
\end{equation}
where $\dot{E}_{\rm GW}$ and $\dot{E}_{\rm Acc}$ are the orbit-averaged outgoing energy fluxes due to GW radiation reaction and accretion-induced drag (that depends linearly on $\xi$, see~\cref{eqn:drag_force}) respectively.
For each term, we have included the corresponding perturbation parameters for book-keeping.
Applying~\cref{eqn:e-balance-wrong} gives
\begin{align}
\begin{aligned}
    \left( \dfrac{df_s}{dt} \right)_E &= \epsilon_{\rm RR} \dfrac{96}{5} \pi^{8/3} G^{5/3} \mathcal{M}_0^{5/3} f_s^{11/3} \\
    &+ \epsilon_{\rm Acc} \dfrac{(-5+6\xi)}{2} \dfrac{f_{\rm Edd}}{\tau} f_s, \label{eqn:en_balance_fdot}   
\end{aligned}
\end{align} 
where we have restored factors of $G$.
Relative to the Keplerian dynamics, we have only kept $\mathcal{O}(\epsilon_{\rm RR})$ and $\mathcal{O}(\epsilon_{\rm Acc})$ contributions, and neglected higher order $\mathcal{O}(\epsilon_{\rm Acc}\epsilon_{\rm RR})$, $\mathcal{O}(\epsilon_{\rm RR}^2)$, and $\mathcal{O}(\epsilon_{\rm Acc}^2)$ terms because we are effectively performing a leading order expansion in a multiple-timescale analysis~\cite{benderorszag} (see also~\cite{Will:2013cza,Chandramouli:2021kts} for other applications of such an approach).
Thus, it is valid to evaluate~\cref{eqn:en_balance_fdot} with the initial value of the masses.
On the other hand, the angular momentum balance law is given by
\begin{align}
 \dot{L} = -\epsilon_{\rm RR} \dot{L}_{\rm GW} - \epsilon_{\rm Acc} \dot{L}_{\rm Acc},\label{eqn:l-balance}   
\end{align}
where $\dot{L}_{\rm GW}$ and $\dot{L}_{\rm Acc}$ are the orbit-averaged outgoing angular momentum fluxes due to GW radiation reaction and accretion-induced drag (that depends linearly on $\xi$, see~\cref{eqn:drag_force}) respectively.
Applying~\cref{eqn:l-balance} results in
\begin{align}
\begin{aligned}
    \left( \dfrac{df_s}{dt} \right)_L &= \epsilon_{\rm RR} \dfrac{96}{5} \pi^{8/3} G^{5/3} \mathcal{M}_0^{5/3} f_s^{11/3} \\
    &+ \epsilon_{\rm Acc} (5+3\xi) \dfrac{f_{\rm Edd}}{\tau} f_s,
    \label{eqn:ang_balance_fdot}    
\end{aligned}
\end{align}
where again the masses are evaluated with their initial values for the reasons mentioned above.
We see that $(df_s/dt)_E - (df_s/dt)_L = -(15/2) f_{\rm Edd}/\tau f_s$, showing that the balance laws do not agree.

Since the inconsistency does not depend on $\epsilon_{\rm RR}$ or $\xi$, the issue does not stem from the additional dissipative fluxes, but simply from the time-dependent nature of the masses. 
To isolate this, we send $\epsilon_{\rm RR} \rightarrow 0, \xi \rightarrow 0$ (or equivalently $\dot{E}_{\rm GW} \rightarrow 0, \dot{E}_{\rm Acc} \rightarrow 0$). 
As noted in~\cite{Caputo:2020irr}, with time-dependent masses, the binding energy $E_b$ is not conserved because the Hamiltonian $\mathcal{H}$ is explicitly time-dependent. 
However, provided that masses change slowly and perturbatively with $\epsilon_{\rm Acc} \ll 1$, the azimuthal and radial actions are adiabatically conserved~\cite{Caputo:2020irr,landau1976mechanics}.
Note that the azimuthal action is simply the angular momentum while the radial action is the integrated (over one orbit) radial momentum.
These imply that angular momentum and eccentricity are conserved adiabatically~\cite{landau1976mechanics}.
For a quasi-circular inspiral, the back-reaction is thus correctly obtained when using angular momentum conservation\footnote{Starting from the Lagrangian with time-dependent masses (no drag), one can derive the equations of motion that directly show that angular momentum is conserved since the Lagrangian admits SO(3) symmetry.}.
The energy balance law that accounts for the adiabatic change in masses is then~\cite{landau1976mechanics} $\dot{E}_b = \epsilon_{\rm Acc} \widetilde{\partial H/\partial t}$, where $\widetilde{ \, \cdot \,}$ denotes orbit-averaging.
When restoring the GW radiation reaction and accretion drag, we obtain the change in binding energy as
\begin{align}
 \dot{E}_b =  \epsilon_{\rm Acc} \widetilde{\dfrac{\partial \mathcal{H}}{\partial t}} - \epsilon_{\rm RR} \dot{E}_{\rm GW} - \epsilon_{\rm Acc} \dot{E}_{\rm Acc}, \label{eqn:Edot_hamiltonian}
\end{align}
which then (for quasi-circular orbits) gives the same back-reaction as when using the angular momentum balance law of~\cref{eqn:l-balance}, resolving the inconsistency.

We note that a similar problem arises in the case of varying-$G$ theories~\cite{Yunes:2009bv,Tahura:2018zuq}, with the Hamiltonian being time-dependent owing to the time-varying $G(t)$. 
When using the incorrect energy balance law $\dot{E}_b = - \epsilon_{\rm RR} \dot{E}_{\rm GW}$ (the equivalent of~\cref{eqn:e-balance-wrong} with no drag-induced flux), this leads to an incorrect back-reaction that propagates to the GW phase, which was obtained by~\cite{Yunes:2009bv}. 
The back-reaction for time-dependent $G(t)$ was subsequently corrected by~\cite{Tahura:2018zuq}, which is consistent with our discussion for accretion (see also~\cite{Caputo:2020irr}).

\section{Additional details and robustness checks of the phenomenological models}
\label{app:template}
In this appendix, we provide complete details on the construction of the phenomenological parametric models, their validity, and tests done with them for Bayesian parameter estimation.

\subsection{RPL model construction}
The RPL model is constructed based on the asymptotic behavior of $\Omega_{\rm GW}(f)$ in the low and high frequency regimes.
The low frequency regime is formally characterized by $f_s \ll f_{\rm turn}$, which is when the energy flux $\dot{E}_{\rm env}$ due to the environment dominates over the GW flux $\dot{E}_{\rm GW}$.
In the high frequency regime when $f_s \gg f_{\rm turn}$, GW emission is instead the dominant energy loss channel. 
In both regimes, we can simplify $dE_{\rm GW}/df$ by performing appropriate asymptotic expansions.

When $f_s \gg f_{\rm turn}$, we perform a ``weak-coupling'' expansion to lowest order in $|\dot{E}_{\rm env}/\dot{E}_{\rm GW}| \ll 1$, which simply results in the vacuum result for $dE_{\rm GW}/df_s$ and $\Omega_{\rm GW}(f)$. 
Thus, we can match $A_{\rm vac}$ to the vacuum amplitude, after accounting for the appropriate frequency scaling of $f^{2/3}$.
Explicitly, we have that
\begin{align}
    A_{\rm vac} = \dfrac{ \pi^{2/3}}{3\rho_c H_0} \iiint dm_1 dq dz \dfrac{R_{\rm GW}(z)p(m_{1},q)}{(1+z)^{4/3} \mathcal{E}(z)} \eta m^{5/3}.
\end{align}

Likewise, when $f_s \ll f_{\rm turn}$, we perform a ``strong-coupling'' expansion to lowest order in $|\dot{E}_{\rm env}/\dot{E}_{\rm GW}| \gg 1$.
Upon scaling out the environmental parameters (since they only enter linearly in $\dot{E}_{\rm env}$), the resulting $dE_{\rm GW}/df_s$ only depends on the source parameters.
We obtain the overall amplitude of the strong-coupling $\Omega_{\rm GW}(f)$ as an integral over the population parameters, after accounting for the appropriate frequency scaling.
In the strong-coupling asymptotic limit, the relative amplitude $A_m$ can then be found using $A_{\rm vac}$ and the overall amplitude of the strong-coupling $\Omega_{\rm GW}(f)$.

We note that~\cref{eqn:RPL_template} is similar to ppE models~\cite{Yunes:2009ke,Maselli:2016ekw}, in the weak coupling limit $f_s \gg f_{\rm turn}$ (or $\alpha \ll 1$), corresponding to the GW dominant regime. 
In this regime, the SGWB model can be used for probing GR deviations in SGWB signals~\cite{Maselli:2016ekw}.
However, in our case, the energy dissipation due to the environmental is not necessarily small compared to GW emisssion, making the ppE model not applicable for typical astrophysical systems.
In fact, the RPL model is effectively a lowest order $[0/1]$ Pad{\'e}-like approximant\footnote{The RPL model is a $[0/1]$ Pad{\'e} approximant in terms of $f^{1/3}$ and when excluding log-terms in the rational polynomial ansatz.}, thereby resumming the ppE model of~\cite{Yunes:2009ke,Maselli:2016ekw} to be valid in the $f_s \ll f_{\rm turn}$ regime.
Unlike the broken power-law model used by~\cite{Cannizzaro:2023mgc}, our RPL model is $C^{\infty}$ and smoothly connects the low and high frequency asymptotic regimes.

\subsection{RPLP model construction}
In the intermediate regime $f_s \sim f_{\rm turn}$, we have that $|\dot{E}_{\rm env}| \sim |\dot{E}_{\rm GW}|$. 
Thus in this non-perturbative regime, both the strong and weak coupling asymptotic expansions will break down\footnote{Typically, when matching two different asymptotic expansions, there should be a buffer zone where the expansions overlap~\cite{benderorszag}. However, in our case, the two different expansions are in terms of the same formal expansion parameter, which is the reason there is no overlapping regime, especially not when $f_s \sim f_{\rm turn}$.} and typically lead to divergent series.
Thus, adding higher order terms in the two asymptotic limits and using a higher-order $[m/n]$ Pad{\'e}-like approximant may not necessarily help improve the accuracy of the RPL model in the intermediate regime.
Since we only keep the lowest order terms in our RPL model, we end up overestimating the signal in the intermediate regime.
Higher-order contributions in the two asymptotic limits will flatten the signal, and thus in order to capture this non-perturbative behavior in the intermediate regime, we turn to a phenomenological approach.

Specifically, we include an interpolating function that captures the non-perturbative behavior for $f_s \sim f_{\rm turn}$, and smoothly vanishes in the $f_s \ll f_{\rm turn}$ and $f_s \gg f_{\rm turn}$ limits.
We accomplish this using the RPLP model that includes a Gaussian correction to the denominator of the RPL model, as given in~\cref{eqn:RPLP}.
The Gaussian correction depends on the parameters $A_g (\alpha)$, $f_{\rm peak}(\alpha)$, and $\sigma$.
To motivate the functional form of $f_{\mathrm{peak}}(\alpha)$, we compare it with the turning frequency $f_{\mathrm{turn}}(\alpha)$, defined as the frequency at which the energy dissipation due to environmental effects equals that of GW emission, i.e., $|\dot{E}_{\rm env}| = |\dot{E}_{\rm GW}|$ and as shown in~\cref{eqn:f_turn,eqn:f_turn_acc}.
The quantity $f_{\mathrm{turn}}(\alpha)$ is obtained by numerical evaluation, redshifted from the source frame to the observer frame by dividing by $(1+z)$, and averaged over the population. 
We find that the fitted $f_{\mathrm{peak}}(\alpha)$ and $f_{\mathrm{turn}}(\alpha)$ are both expressed as power laws in $\alpha$, and that they share the same exponent, with amplitudes differing only by  a factor  $\sim 2$.
In addition, we fit the amplitude $A_g(\alpha)$ and determine $\sigma$ using the full width at half maximum (FWHM) method. 
We find that $\sigma$ remains approximately constant across different values of $\alpha$.

\begin{figure}[t]
\centering
\includegraphics[width=\columnwidth]{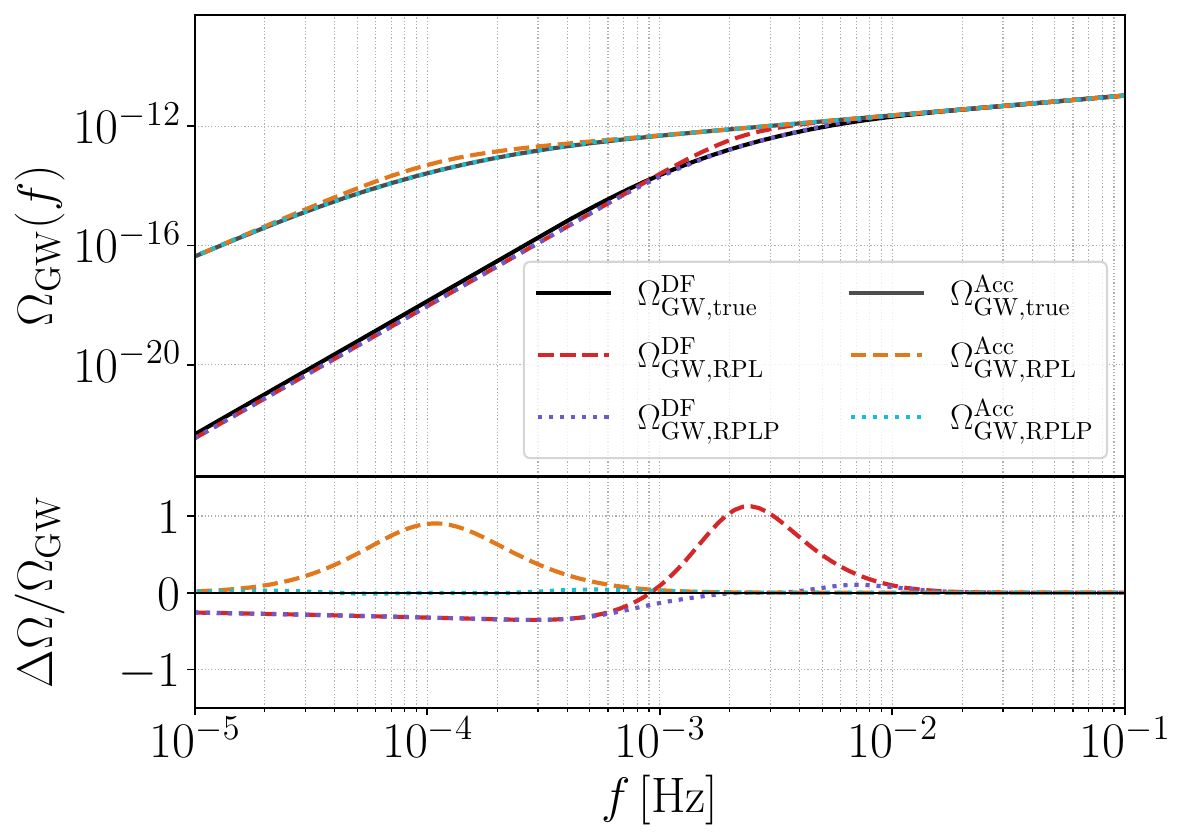}
\caption{
SGWB spectra with dynamical friction ($\rho = 10^{-7}\, \mathrm{g\, cm^{-3}}$) and Eddington-limited accretion ($f_{\mathrm{Edd}} =10 $), shown as solid lines, are compared with two model models: the Rational Power-Law (dashed) and the Rational Power-Law + Peak (dotted).
Lower panels show the residuals $\Delta \Omega = \Omega_{\mathrm{Template}}-\Omega_{\mathrm{GW}}$ for each case.
}
\label{fig:DF_Acc_templates}
\end{figure}
For the cases of dynamical friction and gas accretion, we show a comparison of both the RPL and RPLP models against the true signal in~\cref{fig:DF_Acc_templates}.
Recall that the true signal for dynamical friction and gas accretion are generated using methods described in Sec.~\ref{subsec:eff_DF} and Sec.~\ref{subsec:eff_acc} respectively.
We fiducially set $\rho = 10^{-7} \mathrm{g} \, \mathrm{cm}^{-3}$ for dynamical friction and $f_{\rm Edd} = 10 $ for gas accretion.
The RPL model (blue dashed line) has a significant mismatch with the true signal (black solid line) in the intermediate regime, for both dynamical friction and gas accretion. 
However, the RPLP model (dashed red line) effectively removes this mismatch at intermediate frequencies.

In the case of dynamical friction, a small mismatch arises at low frequencies due to neglecting the $(1+z)$ factor in $\ln[(1+z)f]$, as shown in~\cref{fig:DF_Acc_templates}. 
We do this simplification to scale out the frequency dependence when computing the asymptotic value of $A_m$, and thus underestimating the signal in the low frequency regime.
Note that for gas accretion, the RPLP model matches the signal well in the low frequency asymptotic regime since there are no $\ln[(1+z)f]$ terms involved. 
Thus for gas accretion, the frequency dependence can be trivially scaled out, resulting in a highly accurate asymptotic estimate of $A_m$.

\subsection{Bayesian parameter estimation checks with the RPLP model}
\begin{figure}[t]
    \centering
    \includegraphics[width=\columnwidth]{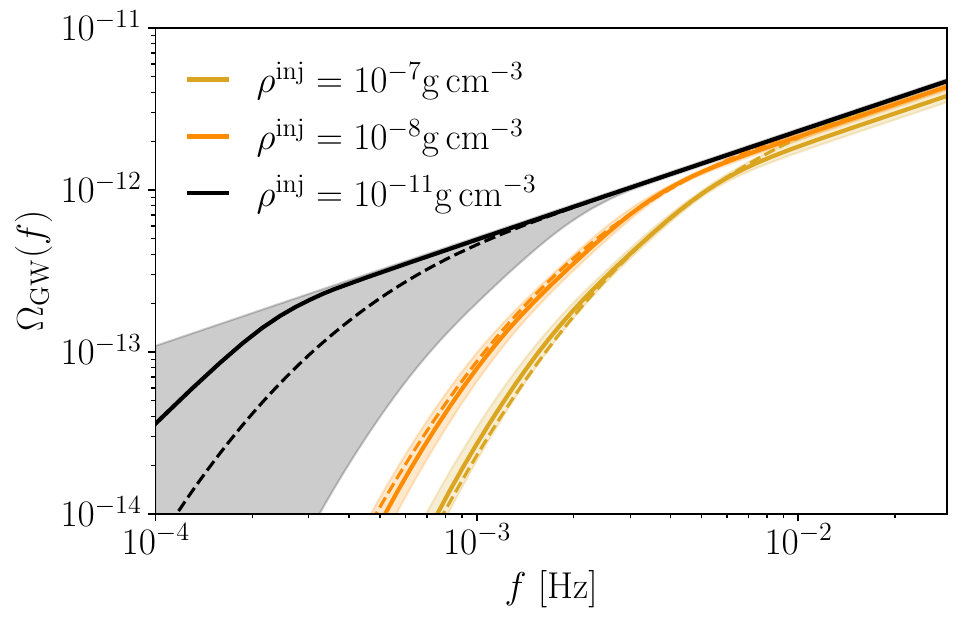}
    \caption{
    Posterior predictives of RPLP model compared against the injected signals (dashed lines) for $\rho^{\rm inj} = \{ 10^{-11}, 10^{-8}, 10^{-7} \}\mathrm{g} \, \mathrm{cm}^{-3}$. 
    For each case, the shaded region corresponds to the $90\%$ credilbe interval, with solid line corresponding to the respective median. 
    Observe that the RPLP model can reconstruct the signal reasonably well within the statistical errors.}    
    \label{fig:DF_sgwb_reconstruction}
\end{figure}

For the case of dynamical friction, since the RPLP model underestimates the true signal at low frequencies, we incurred biases in the marginalized posterior of $\rho$ particularly for large $\rho^{\rm inj}$, as shown in~\cref{fig:DF_template_recovery}.
Notwithstanding these model induced errors, the posterior predictive of the RPLP model agrees reasonably well with the true signal, as we show in~\cref{fig:DF_sgwb_reconstruction}.
Specifically, for the cases of $\rho^{\rm inj} = \{ 10^{-11}, 10^{-8}, 10^{-7} \}\mathrm{g} \, \mathrm{cm}^{-3}$, we compute the posterior predictive in the following way:
for each posterior sample, we generate the RPLP model prediction at each frequency and show the $90\%$ credible interval (shaded region) together with the median (solid line).
In all cases, across frequencies, the posterior predictive agrees well with the injected signal (dashed line) within the $90 \%$ credible interval.
Further, observe that with increasing frequency, the $90 \%$ credible interval of the posterior predictive shrinks due to SNR accumulation.
The main takeaway from the posterior predictive analysis shown here is that although the marginalized posteriors of $\rho$ and $A_{\rm vac}$ display biases, the joint posterior samples describe the signal sufficiently well.

\begin{figure}[!b]
    \centering
    \includegraphics[width=\columnwidth]{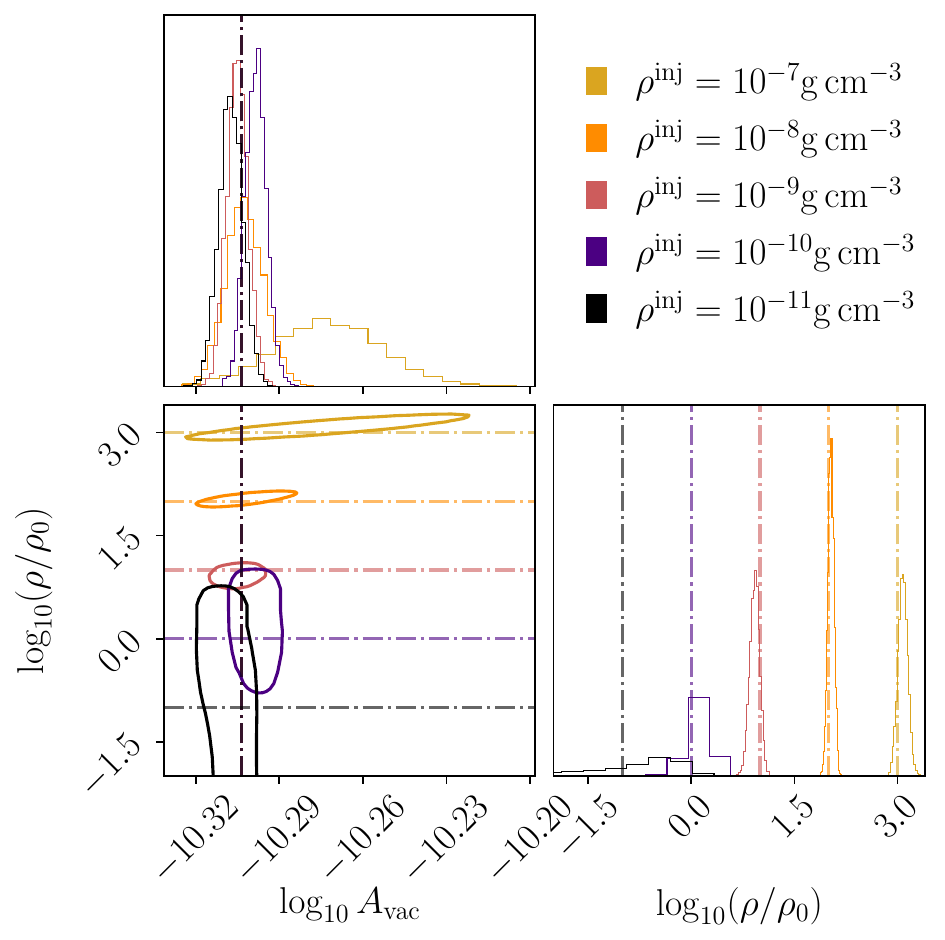}
    \caption{ 
    Marginalized posterior probability of $A_{\rm vac}$ and $\rho$ with mock data generated using the RPLP model. 
    The posteriors for different injected densities are obtained with Galactic foreground parameters fixed.
    Two-dimensional contours correspond to $90\%$ credible regions.
    Dash-dotted lines represent the injected value for $\rho$ and $A_{\rm vac}$.
    Observe that the density is better recovered compared to~\cref{fig:DF_template_recovery}.
    }
    \label{fig:mock_test}
\end{figure}

In~\cref{fig:DF_template_recovery}, we had also found that the marginalized posterior of $A_{\rm vac}$ is consistently biased to smaller values with increasing $\rho^{\rm inj}$.
To isolate this behavior from the bias incurred due to mismatch with the true signal, we performed an injection--recovery check with the RPLP model.
In~\cref{fig:mock_test}, we show the marginalized posteriors of $A_{\rm vac}$ and $\rho$ with the Galactic foreground parameters fixed.
We no longer see any bias in the recovery of $\rho$, implying that indeed the bias we found in~\cref{fig:DF_template_recovery} is mainly due to the mismatch between the RPLP model and the signal (as illustrated in~\cref{fig:DF_Acc_templates}).

We observe that the bias in the marginalized posterior of $A_{\rm vac}$ (relative to its asymptotic vacuum value) in~\cref{fig:mock_test} is smaller than in~\cref{fig:DF_template_recovery}.
Yet, there is still a significant bias in $A_{\rm vac}$ for $\rho^{\rm inj} = 10^{-7} \mathrm{g} \, \mathrm{cm}^{-3}$.
This bias cannot be simply explained by the failure of the RPLP model, since the injection here is with the same model and we do not observe significant biases with $\rho$.
\noindent
We can however explain the $A_{\rm vac}$ bias in the following way.
The asymptotic vacuum value for $A_{\rm vac}$ is meaningful provided that the high frequency asymptotic regime $f_s \gg f_{\rm turn}$ is within the LISA band.
For $\rho^{\rm inj} = 10^{-7} \mathrm{g} \, \mathrm{cm}^{-3}$, $f_{\rm turn}$ shifts to the higher frequencies, thereby pushing the high frequency asymptotic regime to the less sensitive part of the LISA band.
Therefore, using the asymptotic value of $A_{\rm vac}$ as truth to assess bias is not completely valid.
Thus, this contributes to the biases observed in~\cref{fig:DF_template_recovery} for $A_{\rm vac}$.
Put another way, for such large $\rho^{\rm inj}$ the validity regime of the asymptotic matching that is built into the RPLP model affects the inference.

\bibliographystyle{apsrev4-2}
\bibliography{ref.bib}
\end{document}